\documentstyle[10pt,epsf,epsfig,dp_delphititle]{dp_delphi}
\makeindex
\def\DpPaperGroup{EP}
\def\DpPaperRef{98--169}
\def\DpDate{{21 October 1998}}
\def\DpAuthors{DELPHI Collaboration}
\def\DpTitle{{Search for composite and exotic fermions at LEP 2}}
\def\DpSubmit{(Accepted by E. Phys. J. C)}

\begin{document}
\makeatletter
\newcount\@tempcntc
\def\@citex[#1]#2{\if@filesw\immediate\write\@auxout{\string\citation{#2}}\fi
  \@tempcnta\z@\@tempcntb\m@ne\def\@citea{}\@cite{\@for\@citeb:=#2\do
    {\@ifundefined
       {b@\@citeb}{\@citeo\@tempcntb\m@ne\@citea\def\@citea{,}{\bf ?}\@warning
       {Citation `\@citeb' on page \thepage \space undefined}}%
    {\setbox\z@\hbox{\global\@tempcntc0\csname b@\@citeb\endcsname\relax}%
     \ifnum\@tempcntc=\z@ \@citeo\@tempcntb\m@ne
       \@citea\def\@citea{,}\hbox{\csname b@\@citeb\endcsname}%
     \else
      \advance\@tempcntb\@ne
      \ifnum\@tempcntb=\@tempcntc
      \else\advance\@tempcntb\m@ne\@citeo
      \@tempcnta\@tempcntc\@tempcntb\@tempcntc\fi\fi}}\@citeo}{#1}}
\def\@citeo{\ifnum\@tempcnta>\@tempcntb\else\@citea\def\@citea{,}%
  \ifnum\@tempcnta=\@tempcntb\the\@tempcnta\else
   {\advance\@tempcnta\@ne\ifnum\@tempcnta=\@tempcntb \else \def\@citea{--}\fi
    \advance\@tempcnta\m@ne\the\@tempcnta\@citea\the\@tempcntb}\fi\fi}
 
\makeatother
\begin{titlepage}
\pagenumbering{roman}
\CERNpreprint{\DpPaperGroup}{\DpPaperRef} 
\date{{\small\DpDate}} 
\title{\DpTitle} 
\address{\DpAuthors} 
\begin{shortabs} 
\noindent
%

A search for unstable heavy fermions 
with the DELPHI detector at LEP 
is reported. 
Sequential and non-canonical leptons, as well as excited leptons and quarks,
are considered.
The data analysed correspond to an integrated luminosity of 
about 48~pb$^{-1}$ at an $e^+e^-$ centre-of-mass energy of 183~GeV
and about 20~pb$^{-1}$ equally shared between the centre-of-mass 
energies of 172~GeV and 161~GeV. 
The search for pair-produced new leptons establishes 95\% 
confidence level mass limits in the 
region between 70~GeV/$c^2$ and 90~GeV/$c^2$, depending on the channel. 
The search for singly produced excited leptons and quarks establishes 
upper limits
on the ratio of the coupling of the excited fermion
to its mass ($\lambda / m_{f^*}$) as a function of the mass.

\end{shortabs}
\vfill
\begin{center}
\DpSubmit \ 
\end{center}
\vfill
\clearpage
\headsep 10.0pt
\addtolength{\textheight}{10mm}
\addtolength{\footskip}{-5mm}
\begingroup
%
\newcommand{\DpName}[2]{\hbox{#1$^{\ref{#2}}$},\hfill}
\newcommand{\DpNameTwo}[3]{\hbox{#1$^{\ref{#2},\ref{#3}}$},\hfill}
\newcommand{\DpNameThree}[4]{\hbox{#1$^{\ref{#2},\ref{#3},\ref{#4}}$},\hfill}
\newskip\Bigfill \Bigfill = 0pt plus 1000fill
\newcommand{\DpNameLast}[2]{\hbox{#1$^{\ref{#2}}$}\hspace{\Bigfill}}
%
\footnotesize
\noindent
\DpName{P.Abreu}{LIP}
\DpName{W.Adam}{VIENNA}
\DpName{T.Adye}{RAL}
\DpName{P.Adzic}{DEMOKRITOS}
\DpName{T.Aldeweireld}{AIM}
\DpName{G.D.Alekseev}{JINR}
\DpName{R.Alemany}{VALENCIA}
\DpName{T.Allmendinger}{KARLSRUHE}
\DpName{P.P.Allport}{LIVERPOOL}
\DpName{S.Almehed}{LUND}
% ...
\normalsize
\endgroup
\titlefoot{Department of Physics and Astronomy, Iowa State
     University, Ames IA 50011-3160, USA
    \label{AMES}}
\titlefoot{Physics Department, Univ. Instelling Antwerpen,
     Universiteitsplein 1, BE-2610 Wilrijk, Belgium \\
     \indent~~and IIHE, ULB-VUB,
     Pleinlaan 2, BE-1050 Brussels, Belgium \\
     \indent~~and Facult\'e des Sciences,
     Univ. de l'Etat Mons, Av. Maistriau 19, BE-7000 Mons, Belgium
    \label{AIM}}
\titlefoot{Physics Laboratory, University of Athens, Solonos Str.
     104, GR-10680 Athens, Greece
    \label{ATHENS}}
\titlefoot{Department of Physics, University of Bergen,
     All\'egaten 55, NO-5007 Bergen, Norway
    \label{BERGEN}}
\titlefoot{Dipartimento di Fisica, Universit\`a di Bologna and INFN,
     Via Irnerio 46, IT-40126 Bologna, Italy
    \label{BOLOGNA}}
\titlefoot{Centro Brasileiro de Pesquisas F\'{\i}sicas, rua Xavier Sigaud 150,
     BR-22290 Rio de Janeiro, Brazil \\
     \indent~~and Depto. de F\'{\i}sica, Pont. Univ. Cat\'olica,
     C.P. 38071 BR-22453 Rio de Janeiro, Brazil \\
     \indent~~and Inst. de F\'{\i}sica, Univ. Estadual do Rio de Janeiro,
     rua S\~{a}o Francisco Xavier 524, Rio de Janeiro, Brazil
    \label{BRASIL}}
\titlefoot{Comenius University, Faculty of Mathematics and Physics,
     Mlynska Dolina, SK-84215 Bratislava, Slovakia
    \label{BRATISLAVA}}
\titlefoot{Coll\`ege de France, Lab. de Physique Corpusculaire, IN2P3-CNRS,
     FR-75231 Paris Cedex 05, France
    \label{CDF}}
\titlefoot{CERN, CH-1211 Geneva 23, Switzerland
    \label{CERN}}
\titlefoot{Institut de Recherches Subatomiques, IN2P3 - CNRS/ULP - BP20,
     FR-67037 Strasbourg Cedex, France
    \label{CRN}}
\titlefoot{Institute of Nuclear Physics, N.C.S.R. Demokritos,
     P.O. Box 60228, GR-15310 Athens, Greece
    \label{DEMOKRITOS}}
\titlefoot{FZU, Inst. of Phys. of the C.A.S. High Energy Physics Division,
     Na Slovance 2, CZ-180 40, Praha 8, Czech Republic
    \label{FZU}}
\titlefoot{Dipartimento di Fisica, Universit\`a di Genova and INFN,
     Via Dodecaneso 33, IT-16146 Genova, Italy
    \label{GENOVA}}
\titlefoot{Institut des Sciences Nucl\'eaires, IN2P3-CNRS, Universit\'e
     de Grenoble 1, FR-38026 Grenoble Cedex, France
    \label{GRENOBLE}}
\titlefoot{Helsinki Institute of Physics, HIP,
     P.O. Box 9, FI-00014 Helsinki, Finland
    \label{HELSINKI}}
\titlefoot{Joint Institute for Nuclear Research, Dubna, Head Post
     Office, P.O. Box 79, RU-101 000 Moscow, Russian Federation
    \label{JINR}}
\titlefoot{Institut f\"ur Experimentelle Kernphysik,
     Universit\"at Karlsruhe, Postfach 6980, DE-76128 Karlsruhe,
     Germany
    \label{KARLSRUHE}}
\titlefoot{Institute of Nuclear Physics and University of Mining and Metalurgy,
     Ul. Kawiory 26a, PL-30055 Krakow, Poland
    \label{KRAKOW}}
\titlefoot{Universit\'e de Paris-Sud, Lab. de l'Acc\'el\'erateur
     Lin\'eaire, IN2P3-CNRS, B\^{a}t. 200, FR-91405 Orsay Cedex, France
    \label{LAL}}
\titlefoot{School of Physics and Chemistry, University of Lancaster,
     Lancaster LA1 4YB, UK
    \label{LANCASTER}}
\titlefoot{LIP, IST, FCUL - Av. Elias Garcia, 14-$1^{o}$,
     PT-1000 Lisboa Codex, Portugal
    \label{LIP}}
\titlefoot{Department of Physics, University of Liverpool, P.O.
     Box 147, Liverpool L69 3BX, UK
    \label{LIVERPOOL}}
\titlefoot{LPNHE, IN2P3-CNRS, Univ.~Paris VI et VII, Tour 33 (RdC),
     4 place Jussieu, FR-75252 Paris Cedex 05, France
    \label{LPNHE}}
\titlefoot{Department of Physics, University of Lund,
     S\"olvegatan 14, SE-223 63 Lund, Sweden
    \label{LUND}}
\titlefoot{Universit\'e Claude Bernard de Lyon, IPNL, IN2P3-CNRS,
     FR-69622 Villeurbanne Cedex, France
    \label{LYON}}
\titlefoot{Univ. d'Aix - Marseille II - CPP, IN2P3-CNRS,
     FR-13288 Marseille Cedex 09, France
    \label{MARSEILLE}}
\titlefoot{Dipartimento di Fisica, Universit\`a di Milano and INFN,
     Via Celoria 16, IT-20133 Milan, Italy
    \label{MILANO}}
\titlefoot{Niels Bohr Institute, Blegdamsvej 17,
     DK-2100 Copenhagen {\O}, Denmark
    \label{NBI}}
\titlefoot{NC, Nuclear Centre of MFF, Charles University, Areal MFF,
     V Holesovickach 2, CZ-180 00, Praha 8, Czech Republic
    \label{NC}}
\titlefoot{NIKHEF, Postbus 41882, NL-1009 DB
     Amsterdam, The Netherlands
    \label{NIKHEF}}
\titlefoot{National Technical University, Physics Department,
     Zografou Campus, GR-15773 Athens, Greece
    \label{NTU-ATHENS}}
\titlefoot{Physics Department, University of Oslo, Blindern,
     NO-1000 Oslo 3, Norway
    \label{OSLO}}
\titlefoot{Dpto. Fisica, Univ. Oviedo, Avda. Calvo Sotelo
     s/n, ES-33007 Oviedo, Spain
    \label{OVIEDO}}
\titlefoot{Department of Physics, University of Oxford,
     Keble Road, Oxford OX1 3RH, UK
    \label{OXFORD}}
\titlefoot{Dipartimento di Fisica, Universit\`a di Padova and
     INFN, Via Marzolo 8, IT-35131 Padua, Italy
    \label{PADOVA}}
\titlefoot{Rutherford Appleton Laboratory, Chilton, Didcot
     OX11 OQX, UK
    \label{RAL}}
\titlefoot{Dipartimento di Fisica, Universit\`a di Roma II and
     INFN, Tor Vergata, IT-00173 Rome, Italy
    \label{ROMA2}}
\titlefoot{Dipartimento di Fisica, Universit\`a di Roma III and
     INFN, Via della Vasca Navale 84, IT-00146 Rome, Italy
    \label{ROMA3}}
\titlefoot{DAPNIA/Service de Physique des Particules,
     CEA-Saclay, FR-91191 Gif-sur-Yvette Cedex, France
    \label{SACLAY}}
\titlefoot{Instituto de Fisica de Cantabria (CSIC-UC), Avda.
     los Castros s/n, ES-39006 Santander, Spain
    \label{SANTANDER}}
\titlefoot{Dipartimento di Fisica, Universit\`a degli Studi di Roma
     La Sapienza, Piazzale Aldo Moro 2, IT-00185 Rome, Italy
    \label{SAPIENZA}}
\titlefoot{Inst. for High Energy Physics, Serpukov
     P.O. Box 35, Protvino, (Moscow Region), Russian Federation
    \label{SERPUKHOV}}
\titlefoot{J. Stefan Institute, Jamova 39, SI-1000 Ljubljana, Slovenia
     and Department of Astroparticle Physics, School of\\
     \indent~~Environmental Sciences, Kostanjeviska 16a, Nova Gorica,
     SI-5000 Slovenia, \\
     \indent~~and Department of Physics, University of Ljubljana,
     SI-1000 Ljubljana, Slovenia
    \label{SLOVENIJA}}
\titlefoot{Fysikum, Stockholm University,
     Box 6730, SE-113 85 Stockholm, Sweden
    \label{STOCKHOLM}}
\titlefoot{Dipartimento di Fisica Sperimentale, Universit\`a di
     Torino and INFN, Via P. Giuria 1, IT-10125 Turin, Italy
    \label{TORINO}}
\titlefoot{Dipartimento di Fisica, Universit\`a di Trieste and
     INFN, Via A. Valerio 2, IT-34127 Trieste, Italy \\
     \indent~~and Istituto di Fisica, Universit\`a di Udine,
     IT-33100 Udine, Italy
    \label{TU}}
\titlefoot{Univ. Federal do Rio de Janeiro, C.P. 68528
     Cidade Univ., Ilha do Fund\~ao
     BR-21945-970 Rio de Janeiro, Brazil
    \label{UFRJ}}
\titlefoot{Department of Radiation Sciences, University of
     Uppsala, P.O. Box 535, SE-751 21 Uppsala, Sweden
    \label{UPPSALA}}
\titlefoot{IFIC, Valencia-CSIC, and D.F.A.M.N., U. de Valencia,
     Avda. Dr. Moliner 50, ES-46100 Burjassot (Valencia), Spain
    \label{VALENCIA}}
\titlefoot{Institut f\"ur Hochenergiephysik, \"Osterr. Akad.
     d. Wissensch., Nikolsdorfergasse 18, AT-1050 Vienna, Austria
    \label{VIENNA}}
\titlefoot{Inst. Nuclear Studies and University of Warsaw, Ul.
     Hoza 69, PL-00681 Warsaw, Poland
    \label{WARSZAWA}}
\titlefoot{Fachbereich Physik, University of Wuppertal, Postfach
     100 127, DE-42097 Wuppertal, Germany
    \label{WUPPERTAL}}
\titlefoot{On leave of absence from IHEP Serpukhov
    \label{MILAN-SERPOU}}
\titlefoot{Now at University of Florida
    \label{FLORIDA}}
\addtolength{\textheight}{-10mm}
\addtolength{\footskip}{5mm}
\clearpage
\headsep 30.0pt
\end{titlepage}
%
\pagestyle{plain}
\pagenumbering{arabic} 
\setcounter{footnote}{0} %
\large
%
\def\pslash{\not{\hbox{\kern-2.3pt$p$}}}
\def\thslash{\not{\hbox{\kern-2.3pt$\theta$}}}
\def\eslash{\not{\hbox{\kern-2.3pt$E$}}}

\section{Introduction}

It is widely believed that the Standard Model (SM), although extremely
successful at the present energy scale, is not the final theory.
Many possible extensions of the SM discussed in the literature \cite{lit}
predict the existence of new fermions.

This paper reports a search for unstable exotic and excited leptons 
and for excited quarks in DELPHI at
centre-of-mass energies, $\sqrt{s}$, of 183 GeV, 172 GeV and 161 GeV.
Partial results published by DELPHI at $\sqrt{s}=$~161~GeV 
can be found in \cite{delphi1}.
The statistics correspond to an integrated
luminosity of 47.7 pb$^{-1}$ at $\sqrt{s}=183$~GeV,
10 pb$^{-1}$ at $\sqrt{s}=172$~GeV and 10 pb$^{-1}$ at $\sqrt{s}=161$~GeV.

The exotic leptons examined here belong to two classes: sequential
and non-canonical. 
Sequential leptons have gauge quantum numbers
identical to the SM leptons (as for instance the hypothetical 
heavy fourth-generation leptons \cite{bhatta}) 
while non-canonical leptons \cite{djouadi}
have Left-Handed (LH) and Right-Handed (RH) components transforming  
differently from those of SM leptons\footnote{The designation exotic leptons 
is, for some authors, equivalent to non-canonical leptons, while
for others, as in this paper, it encompasses both sequential and  
non-canonical leptons.}. 
Two types of non-canonical leptons are searched for:
mirror leptons which
have the opposite chiral properties of SM leptons, and vector
leptons which have both LH and RH components as isodoublets.
The production and decay modes of sequential and non-canonical
leptons are discussed in sections~\ref{sec:theo1} and \ref{sec:theo2} 
below.


Excited fermions ($f^*$) are expected in models with substructure in the 
fermionic sector. Following the simplest 
phenomenological models \cite{hagiboud}, excited fermions are assumed 
to have both spin and isospin 1/2 and to have both their LH and RH 
components in weak isodoublets (vector-like).
Form factors and anomalous magnetic moments of excited leptons are not
considered in the present analysis.
The production and decay modes of excited
leptons and quarks are discussed in section~\ref{sec:theo3}.

Previous limits set by DELPHI and by other experiments can be
found in references \cite{delphi1,delphi2,delphip3} 
and \cite{others} respectively.

This paper is organized as follows. In section 2 the production and
decay mechanisms of excited and exotic fermions (within the considered
models) are discussed. In section 3 the DELPHI detector and the
used data samples are briefly described. The event selection and
topological classification are discussed in section 4, and the
results are presented in sections 5 and 6.

\section{Production and decay of unstable new fermions}
\label{sec:theo}

The new fermions considered in this paper couple
to the photon and/or to the $W/Z$ gauge 
bosons, according to their 
internal quantum numbers 
and thus could be pair-produced at LEP. 
Single production in
association with  their SM partners is also possible 
but its rate depends on the $ff^*V$ couplings, where
$V$ is a generic gauge boson ($V=\gamma,W,Z$) \cite{djouadi}. 
Excited fermion masses up to $\sqrt{s}$ can be probed 
through single production, depending on the scale $\Lambda$ of the
substructure (which determines the coupling).
In the case of excited quarks only the single production modes
will be considered.
For exotic leptons single production occurs 
because of their mixing with SM leptons. The 
mixing angles, severely restricted by data taken at
LEP1 and in several low-energy experiments (in particular by
the experimental absence of flavour-changing neutral currents), are
constrained to be smaller than $\cal O$~($10^{-1}$) \cite{london}. Given these 
limits and the present luminosities, exotic single lepton production 
is not relevant in most scenarios and will not be considered in the 
present paper.
    
In this paper new fermions are assumed to decay promptly  (decay 
length shorter than about 1 cm). This constraint implies 
mixing angles greater than $\cal O$~($10^{-5}$)
for exotic leptons.
The mean lifetime of excited fermions with masses above 20 GeV/$c^2$
is predicted to be less than 10$^{-15}$ seconds in all 
the cases studied.

\subsection{Sequential leptons}
\label{sec:theo1}

In $ e^+ e^-$ collisions the pair production of heavy  
sequential leptons could proceed through $s$-channel
$\gamma$ and $Z$ exchange for charged leptons ($L^+ L^-$), 
while for neutral leptons ($L^0 \overline{L^0}$) the 
$\gamma$ channel is absent. There is
a $t$-channel W exchange diagram 
for $L^0 \overline{L^0}$ 
which can be neglected, since this contribution involves
the suppressed mixing with the first generation.

The cross-sections, given in reference \cite{bhatta}, 
are essentially the SM cross-section for the second and third generations
reduced by phase-space factors that are functions of the heavy lepton mass
and of the lepton type.

Charged heavy leptons would decay through mixing
into one of the lighter neutrinos or charged leptons and a $W^*$ or a $Z^*$
(for heavy lepton masses above $m_W$ and $m_Z$, the $W$ or $Z$ will be 
on-shell): 
$L^{-} \rightarrow \nu_\ell ~ W^{*-}$ or
$L^{-} \rightarrow \ell^{-} ~ Z^{*}$,
\footnote{In all cases the corresponding decays of the antiparticles are
also implied.} 
where $\ell=e,\mu,\tau$. 

In a similar way, neutral heavy leptons would be allowed to decay through 
mixing into a SM charged lepton or neutrino
and a $W^*$ or a $Z^*$:
$L^{0} \rightarrow \ell^- ~ W^{*+}$
or
$L^{0} \rightarrow \nu_\ell ~ Z^{*}$,
where again $\ell=e,\mu,\tau$. 

The decays into a $W$ boson are largely dominant at the presently accessible
masses ($m \sim m_W  < m_Z $) and are the only ones taken into account.

Cascade decays involving  $L^-$ and $L^0$ 
were not considered, as in any circumstances the lower mass heavy
lepton should instead be detected in the corresponding direct production
reaction. 

Sequential neutrinos are assumed to be Dirac neutrinos
in this analysis.

\subsection{Non-canonical leptons}
\label{sec:theo2}

The non-canonical leptons considered in this paper (mirror and vector
leptons) have the same electrical charge as,
but different weak isospin from the 
corresponding SM leptons. 
Their pair production in $ e^+ e^-$ collisions is thus similar 
to that of the sequential leptons discussed above but with different
vector and axial couplings to the Z. Cross-sections are given in 
reference \cite{djouadi}.

 These new leptons mix with the SM leptons but the non-diagonal terms 
are negligible \cite{london}. They would decay into massive gauge bosons 
plus their ordinary light partner.  
 The decay modes of charged mirror and vector leptons ($E_\ell^{\pm}$) are: 
$E_\ell^{-} \rightarrow \nu_\ell ~ W^{*-}$;
$E_\ell^{-} \rightarrow \ell^- ~ Z^{*0}$. 
 The decay modes of neutral mirror and vector leptons ($N_\ell$) are: 
$N_\ell \rightarrow \ell^- ~ W^{*+}$;
$N_\ell \rightarrow \nu_\ell ~Z^{*0}$.

The decays into a $Z$ boson have a low BR at the presently accessible
masses ($m \sim m_W $) and will not be considered.

As for sequential leptons, 
cascade decays involving  $E^-$  and $N$ 
are also not taken into account.

\subsection{Excited fermions}
\label{sec:theo3}

Pair production of charged excited fermions could proceed via  $s$-channel 
$\gamma$ and $Z$ exchanges in $e^+ e^-$ collisions, 
while for excited neutrinos only 
$Z$ exchange contributes. 
Although   $t$-channel  contributions are also
possible, they correspond to double de-excitation,
and give a negligible contribution to the overall production 
cross-section~\cite{hagiboud}. 

In the single production mode, excited fermions could result
from the  $s$-channel $\gamma$ and $Z$ exchange.
Important additional contributions from t-channel $\gamma$ and
$Z$ exchange arise for excited electron production, while 
t-channel $W$ exchange can be important
for the excited electronic neutrino~\cite{hagiboud}.
For  the  $t$-channel production process,  
the unexcited beam particle is emitted preferentially at low polar
angle and often goes undetected in the beam pipe. 
 
The effective electroweak Lagrangian \cite{hagiboud} associated 
with magnetic transitions 
from excited fermions ${\bf f}^*$ to ordinary fermions ${\bf f}$ has the form 

$$
 L_{{\bf f f}^*} = 
                   \frac{1}{2\Lambda} \overline{{\rm \bf f}^*} 
                    \sigma^{\mu \nu} 
                   \left[ 
                         g f \frac{\bf \tau}{2}  {\bf W}_{\mu \nu} +
                         g^{'}f^{'} \frac{Y}{2} B_{\mu \nu} +
                         g_s f_s \frac{\bf \lambda}{2} {\bf G_{\mu\nu}} 
                   \right] {\rm \bf f}_L       
                   + h.c.
$$
where $\Lambda$ corresponds to the compositeness mass scale, the subscript
L stands for left-handed, $g$, $g^{'}$ and
$g_s$ are the SM gauge coupling constants
and the factors $f$,  $f^{'}$ and $f_s$ 
are weight factors associated with the three gauge groups 
(SU(2) $\times$ U(1) $\times$ SU(3)).
The meaning of these couplings and a more extensive discussion of 
the effective Lagrangian can be found in~\cite{hagiboud}.            
With the assumption $|f|=|f^{'}|=|f_s|$, or assuming that only one of 
the constants $f$ is non-negligible, the cross-section depends 
simply on the 
parameter $f/\Lambda$, which is related to the excited fermion mass 
according to $f/\Lambda = \sqrt{2} \lambda/m_{f^*}$, where $\lambda$ 
is the coupling of the excited fermion. 
 
Excited fermions can decay by radiating a 
$\gamma$,  $Z$  or  $W$. 
For excited quarks, the gluon radiation transition is also possible,
becoming in general the most important decay mode. 
The decay branching ratios are functions
of the $f$, $f^{'}$ and $f_{s}$ coupling parameters of the model.  
Table \ref{tab:lbran} shows
the excited leptons' decay branching ratios for some relevant values of $f$ and
$f^{'}$, and for chosen excited lepton masses.

\begin{table}[hbt]
\begin{center}
{
\begin{tabular}{|l||r|r||r|r|}
\hline
Decay   & \multicolumn{2}{c||} {M=80 GeV/$c^2$} 
& \multicolumn{2}{c|} {M=170 GeV/$c^2$} \\
\cline{2-5}
Channel & {$f=f^{'}$} & {$f=-f^{'}$} & {$f=f^{'}$} & {$f=-f^{'}$} \\
\hline
\hline
$\ell^* \rightarrow \ell\gamma    $ 
& \phantom{a} 100 & 0 & 37 & 0 \\ \hline
$\ell^* \rightarrow \ell Z$ & 0   & 0 & 9  & 36  \\\hline
$\ell^* \rightarrow \nu W      $ & 0   & 100 & 54 & 64  \\\hline
\hline
$\nu^* \rightarrow \nu \gamma   $ & 0 & 100  &  0 & 37  \\ \hline
$\nu^* \rightarrow \nu Z$ & 0 & 0  & 36 & 9  \\ \hline
$\nu^* \rightarrow \ell W$    &  100  & 0  & 64  & 54  \\
\hline
\end{tabular}
}
\end{center}
\caption[]{{Predicted branching ratios in \% for excited lepton decays
(upper part for excited charged leptons, lower part for excited neutrinos).}}
\label{tab:lbran}
\end{table}

For charged excited leptons, the electromagnetic radiative decay is forbidden
if \mbox{$f=-f^{'}$}, and the decay then proceeds 
through the $Z$ and $W$ bosons.  
However, if \mbox{$f=+f^{'}$}, the electromagnetic radiative decay 
branching ratio is close to $100$\% for $m_{\ell^*}$ below 
$m_W$. It decreases above the $W$ threshold, reaching a value of
37\% for $m_{\ell^*} = 170$ GeV/$c^2$.

For excited neutrinos, the situation is reversed, so that 
the electromagnetic partial decay width is zero if $f=+f^{'}$.
However, there is a significant contribution to the 
total decay width from the electromagnetic 
radiative decay if $f \neq f^{'}$, even
if the difference  $f-f^{'}$ is much smaller than $f$ itself. 

In the case of excited quarks, the gluon radiation 
decay mode in general accounts for more than 80\% of the visible width.

The process $e^+ e^- \rightarrow \gamma\gamma(\gamma)$ can be used to
probe compositeness at LEP and thus complement the excited
electron direct searches for the mass region above the kinematical 
threshold. 
In fact, the contribution of the diagram mediated by a virtual excited 
electron to the $\gamma\gamma$ production cross section would lead to a 
modification of the angular distribution.
This effect depends on the excited electron mass $m_{e^*}$ and on the
$e e^* \gamma$ coupling, $\lambda$. 

\subsection{Final state topologies}
\label{sec:theo4}

Many topologies could result from the decay of unstable heavy fermions. 
The different possible production and decay modes
are schematically shown in table~\ref{tab:scheme}. 
The possible final states involve isolated leptons, 
isolated photons, jets, missing energy and missing momentum. 

\begin{table}[hbt]



$$
\begin{array}{|cccc|cccccc|cccccc|}
\hline

\ell & \ell^*     &      &        &      
\ell & \ell^*     &      &        &        &     &  
\ell & \ell^*     &      &        &        &     \\

     & \hookrightarrow & \ell & \gamma &           &
        \hookrightarrow & \nu  & W      &          &      &  &   
        \hookrightarrow & \ell & Z      &      &       \\

    &     &    &    &    &     &   & \hookrightarrow & j & j  & 
    &     &    &  \hookrightarrow  & j & j \\

  &   &    &    &    &    &     &                & \ell & \nu  &  &
    &     &                     & \ell & \ell \\

  &   &    &    &    &    &    &   &  &  &   &  &
    &     &  \nu & \nu \\
\hline

\nu  & \nu^*     &      &        &   
\nu  & \nu^*     &      &        &        &     &  
\nu  & \nu^*     &      &        &        &     \\

     & \hookrightarrow & \nu  & \gamma &      &       
        \hookrightarrow & \ell & W      &      &          &      &    
        \hookrightarrow & \nu  & Z      &      &       \\

    &     &    &    &    &    &   & \hookrightarrow & j & j  & 
    &     &    &  \hookrightarrow  & j & j \\

  &   &    &    &    &    &    &   &   \ell & \nu  & 
    &     &    &                   & \ell & \ell \\
\hline
%
%
q  & q^*     &      &        &    
q  & q^*     &      &        &
 & &&&&&&\\

     & \hookrightarrow & q  & \gamma &      &       
        \hookrightarrow & q & g     &
 & &&&&&&\\
\hline
\end{array}
$$

$$
\begin{array}{|cccccc|ccccccc|}
\hline

 & & \ell^* & \ell^* &  & &  
 & & L^+ (E) & L^-(E) &  & & \\  

\gamma & \ell & \hookleftarrow & \hookrightarrow & \ell & \gamma & 
W & \nu & \hookleftarrow & \hookrightarrow & \nu & W &\\ 

W & \nu & \hookleftarrow & \hookrightarrow & \nu & W & 
 &  &  &  &  &  &\\ 
\hline


 & & \nu^* & \overline{\nu}^* &  & &   
 & & L^0(N) & \overline{L}^0(N) &  & & \\

\gamma & \nu & \hookleftarrow & \hookrightarrow & \nu & \gamma & 
W & \ell & \hookleftarrow & \hookrightarrow & \ell & W &\\

W & \ell & \hookleftarrow & \hookrightarrow & \ell & W 
 &  &  &  &  &  & &
\\
\hline
\end{array}
$$


\caption{Production and decay modes of heavy fermions considered in this
analysis. The upper diagrams correspond to  single production of excited
leptons ($\ell^*,\nu^*$) and quarks ($q^*$), and the lower diagrams to pair
production of excited leptons ($\ell^*,\nu^*$), sequential leptons
($L^{\pm},L^0$), and non-canonical leptons ($E_i,N_i$). The decay products
are charged and neutral leptons ($\ell,\nu$), photons ($\gamma$),
jets ($j$) and gauge bosons ($\gamma,W,Z,g$).}
\label{tab:scheme}
\end{table}

In this analysis, the topologies are classified as $leptonic$ if they 
result from radiative decays of the heavy leptons or from decays into 
W or Z bosons that decay exclusively 
into leptons, and are classified as $hadronic$ 
otherwise. 

Events can be characterized by the number of jets and the number of 
isolated leptons and photons as defined by the reconstruction. 
The different topologies will be referred to 
as $xijk$ according to the following rule: $x$ is $h$ or $\ell$ for 
$hadronic$ or $leptonic$ topologies and $i$ is  the number of jets, 
$j$ is the number of isolated leptons
and $k$ is the number of isolated photons.
As an example, h210 is a hadronic topology
with two jets and one isolated lepton.

\begin{table}[hbt]
\begin{center}
{
\begin{tabular}{|l|c|c|}
\hline
      & \multicolumn{2}{c|}{Topologies} 
\\ \cline{2-3}
Channel & Single production & Pair production
\\
\hline
\hline
$L^\pm \rightarrow \nu W$     &      -                  
& h210,h400 
\\ 
$L^0 \rightarrow \ell W$      &       -                 
& h230(h220),h420
\\ 
\hline
\hline
$E_i\rightarrow \nu_i W$      &        -                
& h210,h400 
\\ 
$N_i \rightarrow \ell_i W$    &         -            
& h230(h220),h420
\\ 
\hline
\hline
$\ell^* \rightarrow \ell\gamma$ &  $\ell$201($\ell$101) 
& $\ell$202 \\ 
$\ell^* \rightarrow \nu W$      &  h210(h200),$\ell$200 
& h210,h400 
\\ 
$\ell^* \rightarrow \ell Z$     &  h220(h210),$\ell$400 & -
\\ \hline
$\nu^* \rightarrow \nu \gamma $ &  $\ell$001             
& $\ell$002
\\ 
$\nu^* \rightarrow \ell W$    &  h210(h200),$\ell$200 
& h230(h220),h420
\\ 
$\nu^* \rightarrow \nu Z$     &  h200,$\ell$200          & -
\\ 
\hline
\hline
$q^* \rightarrow q \gamma$ & h201(h101) 
& -  \\ 
$q^* \rightarrow q g $ & h300 
& -  \\ 
\hline
\hline
$e^+ e^- \rightarrow \gamma\gamma$ & \multicolumn{2}{c|}{$\ell$002}\\ 
\hline           
\end{tabular}
}
\end{center}
\caption[]{Observable topologies corresponding to the different production and
decay modes of unstable heavy fermions.}
\label{tab:topo}
\end{table}

The criteria for selecting isolated particles and jet clustering are
explained in section \ref{sec:selec}, both for $hadronic$ and $leptonic$
events. As will be seen, in the case of the $leptonic$ events
all charged particles are included in the jets and the concept of isolated
leptons is not used. 

Table \ref{tab:topo} shows the relevant topologies for the different production 
and decay channels. The topologies in brackets do not correspond
directly to the physical final state but are often the observed ones.
They become particularly important whenever there are particles produced with 
very low momentum or at small angle to the beam.

Only the topologies that will be considered in this analysis
are indicated in the table. 
Thus in the pair production modes with both heavy leptons
decaying into $W$ bosons
the topologies corresponding to the purely leptonic decays of the $WW$ pair
are not considered, due to their low branching ratio.

Single and double photon final states 
($\ell 001$ and $\ell 002$ topologies)
arise in the case of radiatively 
decaying excited neutrinos. For these topologies, the analyses presented
in references \cite{photons} are used.

\section{The DELPHI detector and the data samples}
\label{sec:detect}

A detailed description of the DELPHI detector
and of its performance can be found in \cite{performance}.
This analysis relies both on the charged particle detection
provided by the tracking system 
and on the neutral cluster detection 
provided by the electromagnetic and hadronic calorimeters.

The main tracking detector of DELPHI is the Time Projection Chamber,
which covers the angular range $20^\circ < \theta < 160^\circ$,
where $\theta$ is the polar angle defined with respect to the
beam direction.
Other detectors contributing to the track reconstruction are
the Vertex Detector,
the Inner and Outer Detectors and the Forward Chambers.
The best momentum resolution obtained for 45~GeV/$c$ muons is 
$\sigma(1/p)=0.57\times10^{-3}$ (GeV/$c$)$^{-1}$.  
The VD consists of three cylindrical layers
of silicon strip detectors, each layer covering the full azimuthal angle.

Electromagnetic shower reconstruction is performed in DELPHI using
the barrel 
and the forward electromagnetic calorimeters, including the STIC 
(Small angle TIle Calorimeter), the DELPHI luminosity monitor.
The energy resolutions of the 
barrel and forward electromagnetic calorimeters 
are parameterized 
respectively as $\sigma(E)/E = 0.043 \oplus 0.32/\sqrt{E}$ and 
$\sigma(E)/E = 0.03 \oplus 0.12/\sqrt{E} \oplus 0.11/E$,
where $E$ is expressed in GeV and the symbol `$\oplus$' implies addition in 
quadrature.
The hadron calorimeter covers both the barrel and forward regions.
It has an energy resolution of
$\sigma(E)/E = 0.21 \oplus 1.12/\sqrt{E}$ in the barrel.

Photon detection in the region between the barrel and the forward
electromagnetic calorimeters (polar angles around 40$^\circ$ and
140$^\circ$) is achieved using the information of a set of lead/scintillator 
counters (40$^\circ$ taggers). The efficiency of the taggers was checked
with Bhabha events and found to be greater than $95\%$.

Finally, muons are identified by their penetration through the iron yoke
of the hadron calorimeter to drift chambers covering both
the barrel and the forward region of the detector. The barrel region is
equipped with three layers of drift chambers, while the end caps contain
two planes. One surrounding layer of streamer tubes completes the coverage
between the two regions.

The effects of experimental resolution, both on the signals and on 
backgrounds, were studied by generating Monte Carlo 
events  for the possible signals and for the SM processes, 
and passing them through the 
full DELPHI simulation and reconstruction chain.
Bhabha events were generated with the Berends, Hollik and Kleiss generator 
\cite{babamc}, while  $e^+ e^- \rightarrow Z \gamma$ events
were generated with PYTHIA \cite{jetset} and KORALZ \cite{koralz}.
PYTHIA was also used for the following processes:
$e^+ e^- \rightarrow WW$, 
$e^+ e^- \rightarrow  W e \nu$,
$e^+ e^- \rightarrow Z Z $,
and $e^+ e^- \rightarrow Z e e$.
In all four-fermion channels, studies with the EXCALIBUR generator \cite{excal} 
were also performed.
The two-photon (``$\gamma\gamma$'') physics events were generated according to
the TWOGAM \cite{twogam} generator for quark channels 
and the Berends, Daverveldt and Kleiss generator \cite{berends} for 
the electron,  muon and tau channels, and also for the Quark Parton Model
giving hadrons.
Compton-like final states originating from an $e \gamma$ collision, 
(with a missing electron in the beam pipe), referred to as 
Compton events, were generated according to \cite{compton_g},
and $e^+ e^- \rightarrow \gamma \gamma$ events according to \cite{kleiss}.

Single and pair excited lepton events
and single excited quark events
were generated according to the cross-sections 
defined in~\cite{hagiboud}, involving $\gamma$ and $Z$ exchange.
Pair production of 
sequential leptons and non-canonical leptons was generated according 
to the cross-sections given in \cite{bhatta} and \cite{djouadi}.
The hadronization and decay processes were simulated by JETSET 7.4 
\cite{jetset}.
The initial state radiation effect was included  at the level of
the generator for  single production, while for pair production it was 
taken into account in the total cross-section.
All the expected decay modes were included in the simulation.

\section{Event selection}

The event selection was performed in three stages. In the first level,  
very general selection criteria were applied
and the events were classified according to the topology scheme described 
above. In the second level, differing selection criteria
were applied to each topology. 
Finally, whenever possible, event
flavour tagging was performed, based on the identification of the 
final state leptons and on other (topology dependent) characteristics 
of the event. Details on each selection level are given below.    

\subsection{Basic event selection}
\label{sec:selec}

The basic event selection and classification was as follows.
Charged particles were considered only if they had 
momentum greater than 0.1 GeV/$c$ and 
impact parameters in the transverse plane and in the beam
direction below 4 cm and 10 cm respectively.  
Neutral clusters were defined as energy depositions in the
calorimeters unassociated with charged particle tracks.
All neutrals of energy above 100 MeV were selected.

Visible energy greater than $0.2\sqrt{s}$ 
in the polar angle region  between $20^\circ$ and $160^\circ$ was required, 
including at least one particle with energy greater than 5~GeV.
Energetic visible particles are expected in all the relevant topologies.
Close to the kinematical limit, these particles 
are produced isotropically. 
In this way the ``$\gamma \gamma$'' background was drastically suppressed
since most of the energy in such events is either 
detected at low polar angles or undetected in the beam pipe. 

Events with measured charged or neutral 
particles having energy greater than $\sqrt{s}$ were rejected.
In addition, at least one charged particle in the polar angle region between 
$25^\circ$ and $155^\circ$ with associated hits in the Vertex Detector was
required. This criterion is useful in rejecting cosmic ray background.

Events with at least six detected charged particles 
were selected for the $hadronic$ topologies, and those
with not more than five for the $leptonic$ topologies.

Charged particles were considered isolated if, in 
a double cone centred on their track with internal and external half 
angles of $5^\circ$ and $25^\circ$, the total energy associated to charged 
and neutral particles was below 1 GeV and 2 GeV respectively.
The energy of the particle was redefined as the sum of the energies
contained inside the inner cone. This energy
was required to be greater than 4 GeV.
In all $hadronic$ topologies with isolated leptons, these were required to have
associated hits inside a 2$^\circ$ cone in at least two layers of the 
Vertex Detector.

Isolated charged particles with an associated electromagnetic energy greater 
than 20\% of their measured momentum were loosely identified as electrons.
In $hadronic$ topologies, they were also required to have an associated 
hadronic energy lower than 15\% of their measured momentum.
Isolated particles were classified as muons by requiring an 
electromagnetic energy lower than 20\% of their measured momentum and 
at least one associated hit in the muon chambers.

In both $hadronic$ and $leptonic$ topologies, 
energy clusters in the electromagnetic calorimeters were
considered to be due to photons if there were no tracks associated to them 
and there were  
no hits inside a 2$^\circ$ cone in more than one layer of the 
Vertex Detector and if at least 
90$\%$ of any hadronic energy was deposited in the 
first layer of the hadron calorimeter. 

Photons were considered to be isolated if, in a double cone 
centred on the cluster and having internal and external half angles of
$5^\circ$ and $15^\circ$, the total energy deposited was
less than 1 GeV. 
The energy of the photon was redefined as the sum of the energies
of all the particles inside the inner cone and
no charged particles above 250~MeV/$c$ 
were allowed inside this cone. 
The photon energy had to be greater than 5~GeV.
No recovery of converted photons was attempted.
In all the studied topologies, photons were required to be 
above 10$^\circ$ in polar angle. In addition, for the $leptonic$ 
topologies the most energetic photon in the
event was required to have an energy greater than 10 GeV.

The search for jets in the selected events was performed with
the Durham jet algorithm~\cite{durham}.
In this algorithm, a resolution variable
$$y_{ij} =
  2 \cdot \frac{\min(E_{i}^{2},E_{j}^{2})}{E_{vis}^{2}} \cdot 
 (1-\cos\theta_{ij})$$
is computed for all pairs of particles.
$E_{i,j}$ are the energies of the particles, $\theta_{ij}$ is their opening 
angle, and $E_{vis}$ is the visible energy in the event.
The pair for which $y_{ij}$
is smallest is replaced by
a pseudoparticle with four-momentum equal to the sum of their
four-momenta.  
In this analysis, the algorithm is used in two different ways:

\begin{itemize}
\item 
the procedure is iterated until all pseudo-particle pairs have
$y_{ij}$ larger than a certain $y_{cut}$ value. 
A cut-off value of $y_{cut}$=0.003 was used.
This relatively low value of $y_{cut}$ is well suited for topologies
with many jets.
\item
the procedure is iterated until all particles are clustered into a certain 
pre-defined number of jets (N$_{jets}$). In this case the
value of $y_{cut}$ at the last iteration,
$y_{cut(N_{jets}+1 \rightarrow N_{jets})}$, as
well as $y_{cut(N_{jets} \rightarrow N_{jets}-1)}$,
characterize the event topology.
\end{itemize}

For $hadronic$ events, all neutral and charged particles except
isolated leptons and photons were included in the jets.
The algorithm was applied four times, requiring N$_{jets}$=1, 2, 3 and 4. 
In order to increase the purity of the 2-jet event sample, only events 
with $y_{cut(3 \rightarrow 2)}<$0.06 and $y_{cut(2 \rightarrow 1)}>$0.001
were kept. 
Similarly, for the 3-jet events the $y_{cut}$ variables were constrained
to $y_{cut(3 \rightarrow 2)}>0.003$ and $y_{cut(4 \rightarrow 3)}<0.001$. 

For the $leptonic$ events, only the isolated photons were left out of the 
jets. Charged particles were not treated as isolated objects, but 
clustered into jets referred to as `low multiplicity' jets or 'leptonic'
jets. 
This allows for the fact that taus can decay into 
several charged and neutral particles, and electrons can be accompanied 
by other electrons and photons due to interactions with matter.
In this case the algorithm was applied with $y_{cut}$=0.003.
Whenever the resulting number of jets was lower than the number of 
isolated leptons previously found (N$_{lept}$) the algorithm was 
applied once more imposing N$_{jets}=$N$_{lept}$.

The jets in $leptonic$ events were loosely identified as electrons or muons
according to the criteria described for isolated leptons. For a jet to be 
identified as a muon, it was also required  not to contain 
more than two tracks.

Jets were classified as charged if they 
contained at least one charged particle. 
In the case of the $hadronic$ ($leptonic$) topologies, only events with all 
jets classified as charged and with axes in the polar 
region between 20$^\circ$ (25$^\circ$) and 160$^\circ$ (155$^\circ$) were 
retained.  
 
\subsection{Selected {\em hadronic} events}
\label{sec:had}

\subsubsection{Single production topologies}
\label{sec:hadsing}

In {\em hadronic} events from single heavy lepton production,
the jets originate from the decay of a $W$ or a $Z$
which is not produced at rest.
The candidates must have
two charged jets with high acollinearity ($A^{jj}_{col}$) and acoplanarity
($A^{jj}_{cop}$)
\footnote{ The acoplanarity is defined as the acollinearity in the plane
perpendicular to the beam.},
and the 2-jet system
must have a high mass (M$_{jj}$) and a high momentum ($P_{jj}$).
The main backgrounds for these topologies are 
$e^+ e^- \rightarrow q \overline{q} (\gamma)$ events, including 
radiative returns to the $Z$ ($e^+ e^- \rightarrow Z \gamma$) where 
the photon was lost in the beam pipe, and semileptonic 
decays of $W$ pairs.
In the first case the events are characterized by two acollinear 
jets.  
In addition, radiative return events have a high missing momentum
($\pslash$) at low polar angle ($\thslash$).
The semileptonic $WW$ events are characterized by the fact that the mass 
recoiling against the 2-jet system ($M_{R}$) should be close to the $W$ mass,
which is not true for the signal.
Fully hadronic $WW$ events rejection can be achieved by cutting 
tighter in the event topology variable $y_{cut(3 \rightarrow 2)}$.

\begin{table}[htb]
\begin{center}
{
\begin{tabular}{|c||c|c|c|c|} \hline
         & \multicolumn{4}{c|}{Selection variables} \\ 
\cline{2-5} 
{\bf Topol} & {\bf angles} & {\bf masses} &   
{\bf other criteria}  &~~{\bf fit}~ \\ \hline\hline      
h{\bf 300} & min(A$^{jj})>40^\circ$ & 
           &                                & 4C  \\
\hline
h{\bf 200} & A$^{jj}_{col}>40^\circ$  & M$_{jj} \in [40,100]$~GeV/$c^2$  
           & $y_{cut(3 \rightarrow 2)}<0.01$  & 1C   \\
           & A$^{jj}_{cop}>25^\circ$  & M$_{R} < 60$~GeV/$c^2$ if {\bf W}      
           & $\pslash>0.11~\sqrt{s}$  &    \\    
           & $\thslash>25^\circ$      & M$_{R} < 75$~GeV/$c^2$ if {\bf Z}
           &                          &    \\
\hline           
h{\bf 210} & A$^{jj}_{col}>30^\circ$  & M$_{jj} > 40$~GeV/$c^2$             
           &                          & 1C \\
           & A$^{jj}_{cop}>15^\circ$  & M$_{R} < 60$~GeV/$c^2$ if {\bf W} 
           &                          &  \\ 
           & $\thslash>20^\circ$      & M$_{R} < 75$~GeV/$c^2$ if {\bf Z}
           &                          & \\
\hline 
h{\bf 220} & A$^{jj}_{col}>20^\circ$  &  
           &                          & 5C \\
           & A$^{jj}_{cop}>10^\circ$  & 
           &                          & 1C \\
           & $\thslash>20^\circ$      &
           &                          & \\ 
\hline  
h{\bf 201} & A$_{iso}^{\gamma} > 25^{\circ}$                & 
           & $E_{\gamma} > 20$ GeV                          & 4C \\
           & $\theta_{\gamma}>40^{\circ}$                   &
           &                                                &    \\
\hline
h{\bf 101} & A$^{j\gamma}_{col}<30^\circ$   & 
           & $E_{\gamma} > 30$ GeV          & 4C \\
           & $\theta_{\gamma}>40^{\circ}$  &
           &                                &    \\
\hline
\end{tabular}
}
\end{center}
\caption[]{Selection criteria for single production $hadronic$ topologies.
The notation for the topologies is the one defined in section~\ref{sec:theo4}.}
\label{tab:selhad}
\end{table}

{\em Hadronic} events from single excited quark production
are characterized by the presence of three jets (h300) in the case of the decay
through the gluon and by one energetic photon with a large isolation angle
(A$_{iso}^{\gamma}$) and two jets (h201)
in the case of the electromagnetic radiative decay.
The 3-jet topologies were selected using the event topology variable
$y_{cut(3 \rightarrow 2)}$  and the minimum angle between jets, min($A^{jj}$). 
The photon is expected to have a rather large energy ($E_\gamma$) and 
an isotropic polar angle distribution ($\theta_\gamma$).  
The main backgrounds for these topologies are 
$e^+ e^- \rightarrow q \overline{q} (\gamma)$ events and hadronic decays of $W$
pairs.
Near the kinematic limit the spectator quark is essentially produced at rest
and the observed topologies are then h200 and h101. While the h101 
corresponds to a very clear signature, in the h200 the SM $q \overline{q}$    
events constitute an enormous and irreducible background.

In order to improve the estimation of the momentum and energy of the jets,
a kinematic constrained fit was applied to the selected events.
The constraints imposed depend on the particular final state being studied. 
In several of the relevant $hadronic$ final states, jet pairs come from
the decay of W or Z bosons. Therefore, the invariant mass of the two-jet 
system can be required to be either $m_W$ or $m_Z$. 
Since the h200 and h210 topologies can arise from both the $W$ and the $Z$ 
channels (see tables \ref{tab:scheme} and \ref{tab:topo}) 
the fit was performed twice for these topologies, 
using $m_W$ and $m_Z$.
If there are no undetected particles, energy and momentum conservation can 
be imposed. This is the case for the topologies with isolated photons
and for the h300 topology (4C fit). For the h220 topology the fit 
was performed both requiring only the invariant mass of the two-jet
system to be $m_Z$ (1C) and imposing the additional constraint of 
energy-momentum conservation (5C).
The input quantities for the fit are basically the measured 
energies and momenta of the objects (particles or jets). 
Lagrange multipliers are used to make a constrained fit to both the
energies and the directions of the jets and isolated particles. 
The fit requires a $\chi^2$
value to be optimized, subject to the given constraints on the 
reconstructed final state objects.
The details of the fitting procedure, including the
errors on the input variables, can be found in reference \cite{fit5c}.

The main selection criteria for the different topologies are summarized in
table~\ref{tab:selhad}.
The number of constraints imposed for each topology is also given in the
table. In all cases, only events with a $\chi^2$ per
degree of freedom lower than 5 were retained.
Whenever a topology was relevant for more than one search channel, some of
the selection criteria could vary from one case to the other. This is
indicated in table~\ref{tab:selhad} for the recoil mass cut in the case of
the h200 and h210 topologies.

The selection criteria allow quite an
efficient background rejection. The cut values 
were tuned for each topology in order to optimize the signal to noise ratio.
For the h200 topology, events with signals in more than  
two 40$^\circ$ taggers inside a 20$^\circ$ cone centred on the direction 
of the missing momentum were rejected. This criterion is useful in rejecting 
$q \overline{q} \gamma$ events in which the photon was lost in the region 
between the electromagnetic calorimeters.

In all the topologies with two jets in the final state,
events were required
to have a charged multiplicity of at least eight.

\subsubsection{Pair production topologies}
\label{sec:haddoub}

The relevant topologies for the pair production of charged heavy leptons
can have two jets and one lepton or four jets, resulting from the decay 
of the two $W$s (see table~\ref{tab:scheme}). 
As mentioned, fully leptonic decay modes, with their rather small 
branching ratios, will not be considered in this analysis.
The main topologies are thus h210 and h400.

\begin{table}[htb]
\begin{center}
{
\begin{tabular}{|c|c||c|c|c|c|} \hline
{\bf Heavy} & {\bf Topo-} & \multicolumn{4}{c|}{Selection variables} \\ 
\cline{3-6} 
{\bf lepton} & {\bf logy} & {\bf leptons} &  ${\bf y}_{{\bf cut}}$  
& {\bf missing} & {\bf other criteria} \\ \hline\hline      

{\bf Neutral} & h230 \& &
$N_{lept}\geq 2$ & $y_{cut(2 \rightarrow 1)}>0.03$ 
& $\thslash > 20^{\circ}$  &            \\
       & $\!\!\!\!\!$h420  &  &  &  &  \\
\hline           
{\bf Charged} & h210  &
$N_{lept}>0$ &    & $\thslash>25^\circ$ &  $E^{15}_{mis}<0.5$ \\ 

  &   &    &    & $\eslash > 0.5 \sqrt{s}$  & $A^{iso}_{lept}>50^\circ$     \\ 
  &         &                  &    & $\pslash_T > 10$~GeV/$c$ &    \\ 
\cline{2-6} 
          & h400 &
$N_{lept}=0$ & $y_{cut(3 \rightarrow 2)}>0.03$  & $\thslash>30^\circ$ 
& A$^{WW}<140^\circ$  \\ 

  &   &   & $y_{cut(4 \rightarrow 3)}>0.003$  & $\eslash > 0.3 \sqrt{s}$ & 
$\pslash / \eslash <0.7$    \\ 
\hline
\end{tabular}
}
\end{center}
\caption[]{Selection criteria for pair production $hadronic$ topologies.}
\label{tab:selhad2}
\end{table}

In the case of neutral heavy leptons, two additional charged leptons are
present in the final state, 
and the main topologies are, thus, h230 and h420.
All $hadronic$ events with at least two isolated leptons found 
were considered.

The additional final state leptons present in signal events for  neutral
heavy lepton production constitute a rather clear signature.
On the contrary, for 
the charged heavy leptons search channels there is a nearly irreducible 
background from $WW$ events. 
Signal events are characterized by the presence of two additional neutrinos,
seen as additional missing energy. The missing momentum will in general
be neither forward nor aligned with the directions of the jets.
Thus, missing energy ($\eslash$) and 
transverse missing momentum ($\pslash_T$) are expected and the
energy in a $15^\circ$ cone around the direction of the missing
momentum ($E^{15}_{mis}$) will be low.

In the final state topology with no isolated leptons (h400), $qq(\gamma)$ 
background can be rejected using the $y_{cut}$ variables to select 4-jet
events.
In this 4-jet topology, the two $W$ candidates (i.e. the two jet pairs 
supposed to result from the decay
of the $W$ bosons) were found by trying all the possible combinations
and choosing the one for which the jet-jet invariant masses best reproduce
the $W$ mass. The angle between the two W candidates ($A^{WW}$) is expected
to be close to $180^\circ$ for the $WW$ background and lower for the signal,
due to the presence of the two additional neutrinos. 
In the final state with one isolated lepton (h210), it is also 
required that the lepton is well isolated ($A^{iso}_{lept}$).
The selection criteria applied in the search for both neutral and charged
pair-produced heavy leptons are summarized in table~\ref{tab:selhad2}.

%
%
%
%
%
%
%
  
\subsection{Selected {\em leptonic} events}
\label{sec:lep}

Events classified as {\em leptonic} can originate from radiative decays of
heavy leptons, in which case there will be photons in the final state,
or from decays into $W$ or $Z$ bosons decaying into leptons, in which case 
there will be only leptonic jets involved (see table~\ref{tab:scheme} and 
table~\ref{tab:topo}).
The two analyses are quite different and will be treated separately. 

\subsubsection{{\em Leptonic} events without isolated photons}
\label{sec:lep1}

The topologies considered in this section are $\ell200$, two low multiplicity 
jets only, and $\ell400$, four low multiplicity jets. As mentioned, 
they arise whenever there are $W$ or $Z$ bosons decaying leptonically. 
Since these topologies arise only in single production
modes, they have to be considered in the search for excited leptons
only.

\vspace{0.5cm}
\noindent
{\it $\ell200$ topology \\}
\noindent
The signal events are characterized by the presence of two acoplanar 
leptonic jets 
and missing energy.
The background for this topology comes essentially from 
$e^+ e^- \rightarrow \ell^+ \ell^- (\gamma)$ processes, in particular Bhabha 
events where the photon is lost, and from leptonic decays of $W$ pairs.
In the $WW$ background events, both the leptonic jets 
come from $W$ decays, having
a large momentum. The general selection criteria were the following:

\begin{itemize}
\item
A$_{col}>10^\circ$,
\item
A$_{cop}>10^\circ$,
\item
$\pslash > 0.11 \sqrt{s}$,
\item
$\theta > 30^\circ$ for both leptonic jets.
\end{itemize}

As seen in table \ref{tab:topo}, the topology $\ell200$ can arise
in several different channels. 
In addition to the general selection criteria for the topology, 
a different specific cut was included for each decay channel.
This cut depends on the origin of the
leptonic jets present in the
final state. They can be 
spectator leptons produced together with the heavy one, products of
the decay of the heavy lepton or products of the decay of a $W$ or
$Z$ boson. 

If the decaying excited lepton is charged, 
$\ell \ell^* \rightarrow (\ell \nu W,\ell \ell Z) \rightarrow 
\ell \ell \nu \nu$, 
the lower energy leptonic jet
is expected to be the spectator lepton and  
the momentum of the least energetic charged jet was required
to be lower than 0.11$\sqrt{s}$.

For excited neutrinos,
in the case of the decay via a $W$
($\nu \nu^* \rightarrow \nu \ell W \rightarrow \ell \ell \nu \nu $) 
the mass recoiling against one of the two leptonic jets was required 
to be in the W mass region (70~GeV$/c^2 < m < 110$~GeV$/c^2$),
while in the case of the decays via a $Z$
($\nu \nu^* \rightarrow \nu \nu Z \rightarrow \nu \nu \ell \ell $)
the invariant mass of the two leptonic jets had to be 
between 80 GeV/$c^2$ and 100 GeV/$c^2$.

\vspace{0.5cm}
\noindent
{\it $\ell400$ topology \\}
\noindent
This topology can arise 
in the case of a singly produced charged excited lepton decaying 
via a $Z$ boson.
For signal events, two of the leptonic jets result from the $Z$ 
decay and thus have a large invariant mass, while the second pair of 
leptonic jets
has in general a low invariant mass.

The background for this topology comes from 4-fermion processes.
In this analysis it was required that at least one of the 
leptonic jets had been 
previously identified as an isolated lepton (see section~\ref{sec:selec}) 
to avoid hadronic contamination.

\subsubsection{{\em Leptonic} events with isolated photons}
\label{sec:lep2}

Final states resulting from
singly produced charged excited leptons decaying radiatively are 
characterized by 
an energetic photon (E$_\gamma$) in the central region of the detector
and two 
low multiplicity jets ($\ell$201 topology). 
Near the kinematic limit one of these jets will not be observed due to
its low momentum ($\ell$101 topology).
This final state topology is particularly relevant when the t-channel 
cross-section dominates ($e^*$ single production) and the spectator 
lepton is frequently lost in the beam pipe. 

The main background for the $\ell$201 topology comes from 
$e^+ e^- \rightarrow \ell^+ \ell^-$ radiative events.
Events having a photon emitted at very low polar angle ($\theta_\gamma$) or 
with low energy are 
easily eliminated. However, events with a hard isolated photon constitute an 
irreducible background.

For the $\ell$101 topology, the background comes from 
Bhabha events where one electron was lost or misidentified as a photon, and 
from Compton events. 
In Bhabha events, there are two essentially back-to-back particles
in the forward regions of the detector, i.e., the jet-photon space angle 
(A$^{j \gamma}$) is around 180$^\circ$.
In Compton events the charged jet and the photon are acollinear 
and there is a large amount of energy deposited at relatively low
polar angles. 

Final states resulting from pair-produced charged excited leptons 
decaying radiatively are characterized
by the presence of two leptonic jets and two hard photons in the
detector. 
Possible background comes from doubly radiative 
$e^+ e^- \rightarrow \ell^+ \ell^-$ events.
In signal events there must be two lepton-photon combinations with 
compatible invariant masses, which correspond to the excited lepton
invariant mass. The relevant variable is the minimum
of the lepton-photon invariant mass differences:
$
\Delta m_{\ell \gamma} = min ( 
|m_{\ell1 \gamma1}-m_{\ell2 \gamma2}|, |m_{\ell1 \gamma2}-m_{\ell2 \gamma1}|)
$.

In three- or four-body topologies, 
the energies can be rescaled by imposing energy and momentum conservation 
and using just the polar and azimuthal angles, which are well measured in
the detector. This procedure can significantly improve the energy resolution.
The compatibility of the momenta calculated from the angles with the measured
momenta was quantified on a $\chi^2$ basis.\footnote{The $\chi^2$ parameter was
defined separately for charged jets ($\chi^2_{charged}$) and photons 
($\chi^2_{photons}$) as $ \chi^2~=~\frac{1}{n} \sum_{i=1,n}
  \left(\frac{p_i^{calc}-p_i^{meas}}{\sigma_i}\right)^2$
where $n$ is the number of measured particles, 
$p_i^{meas}$ are the measured momenta or energies and
$p_i^{calc}$ are the momenta calculated from the kinematic
constraints. 
$\sigma_i$, the quadratic sum of the errors on $p_i^{calc}$
and $p_i^{meas}$, is defined in reference \cite{delphip3}}.
For two-body topologies, the same method can be applied assuming the presence
of a third particle along the beam direction.

Only events with $\chi^2<5$ either for photons or for charged particles were
kept. The fact that the condition $\chi^2<5$ was not applied simultaneously
to the photons and to the charged particles allows events in the electron and
muon channels with photons near the boundaries of the calorimeter modules 
(where 
electromagnetic energy can be badly reconstructed) to be kept.
The rescaling procedure can also be applied to the tau channel because
the charged decay products nearly follow the
direction of the primary tau. However, due to the neutrinos, the rescaled 
momenta of the charged jets are expected to be substantially different from 
the measured ones.
This can be used in the tau lepton identification through a cut
$\chi^2_{charged}>5$.
In addition to this criteria, the events selected in 3-body topologies
were checked 
for their coplanarity (the sum of the angles between the particles had to be
greater than 355$^\circ$).

\begin{table}[htb]
\begin{center}
{
\begin{tabular}{|c||c|c|c|} \hline
         & \multicolumn{3}{c|}{Selection variables} \\ \cline{2-4}
{\bf Topology} & {\bf jet variables} & {\bf photon variables} & 
{\bf other}  \\ \hline\hline
                             
{\bf $\ell$201 $e$}  & p$_{jet_1}>10$~GeV/$c$   & $\theta_{\gamma}>40^\circ$ &
                                       \\ \cline{1-1}\cline{3-3}
                    
{\bf $\ell$201 $\mu$}   &                     &                            &
                                       \\
\cline{1-3}

{\bf $\ell$201 $\tau$}  &                    & $\theta_{\gamma}>20^\circ$ &
                                       \\
\hline

{\bf $\ell$101 $e$}     &                     & E$_{\gamma}>0.22~\sqrt{s}$    &
                  $100^\circ<$A$^{j\gamma}<179^\circ$  \\ \cline{1-1}\cline{4-4}
                  
{\bf $\ell$101 $\mu$}   & p$_{jet}>10$~GeV/$c$   & $\theta_{\gamma}>40^\circ$  &
                                            \\ \cline{1-1}\cline{4-4}
                  
{\bf $\ell$101 $\tau$}  &                     &                             &
                  $100^\circ<$A$^{j\gamma}<179^\circ$  \\
%
\hline\hline

{\bf $\ell$202 $e$}  & p$_{jet_1}>10$~GeV/$c$  &   &
  $\Delta m_{\ell \gamma} < 15$~GeV/$c^2$        \\ \cline{1-1}\cline{4-4}
                    
{\bf $\ell$202 $\mu$} & p$_{jet_2}>10$~GeV/$c$ &                      &
  $\Delta m_{\ell \gamma} < 10$~GeV/$c^2$       \\
\cline{1-2}
\cline{4-4}

{\bf $\ell$202 $\tau$}  &                 &   &
  $\Delta m_{\ell \gamma} < 20$~GeV/$c^2$          \\
\hline
\end{tabular}
}
\end{center}
\caption[]{Selection criteria for $leptonic$ topologies with isolated photons.
The topologies below the double line correspond to the pair production.}
\label{tab:sellep}
\end{table}

The backgrounds and efficiencies depend significantly
on the flavour of the final state leptons involved in these topologies. 
Thus, different 
selection criteria are used for the different flavours.
The main selection criteria are summarized in table \ref{tab:sellep}.

For the $\ell101$ topology, an additional cut was applied in the electron
channel: events that have a rescaled momentum greater
than 0.18$\sqrt{s}$
assigned to the particle lost along the beam direction were rejected. This
criterion is useful to eliminate Compton events.


\subsection{Events with photons only}
\label{sec:phot}

Neutral excited leptons can give rise to single or 
double photon events.
Thus, excited neutrinos produced in pairs and decaying radiatively to 
an ordinary neutrino would be tagged through the $\ell$002
topology and those produced singly would be detected through the
$\ell$001 topology.
For these topologies, the analyses described in \cite{photons} are used.

The SM processes constituting the background to both topologies are 
essentially  
QED $e^+ e^- \rightarrow \gamma\gamma$,
$e^+ e^- \rightarrow Z\gamma(\gamma)$ with $Z \rightarrow \nu \nu$,
radiative Bhabhas and Compton events.  

In order to reduce drastically the Bhabha and Compton background, 
single gamma events were required to have polar angles above 
45$^{\circ}$ and no other electromagnetic energy deposition was allowed
out of a 20$^{\circ}$ cone.
Furthermore, it was required that the photons had a line of flight 
compatible within 
15$^{\circ}$ with the shower direction reconstructed by the HPC calorimeter.
This criterion and an additional
selection based on the HCAL were applied to veto cosmics.

The two-photon sample was selected requiring at least two photons 
satisfying the following criteria:
\begin{itemize}
\item Energy greater than 25\% of the centre-of-mass energy 
      in the polar region between 25$^{\circ}$ and 155$^{\circ}$.
\item At least three HPC layers with more than 5\% of the total 
      electromagnetic energy, for HPC energy depositions
      not pointing to the $\phi$ intermodular zones.
\end{itemize}
Furthermore, the hadronic energy was required to be less than 15\% 
of the total 
deposited energy unless the photon fell in the HPC $\phi$ cracks,
in which case it was required that the HCAL first layer energy deposition
was greater than 90\% of the total hadronic energy.

The $\gamma\gamma$ sample was enriched through the recovery of photons 
converted after the VD detector.  
The number of converted photons was limited to one per event and their
recovering was performed in a slightly reduced geometrical acceptance 
($\theta > 30^{\circ}$), in order to keep a high level of background 
rejection.
A converted photon was defined as an energy deposition 
associated to a charged particle track and with no 
VD track elements within 2$^{\circ}$ and 6$^{\circ}$
for the barrel ($\theta > 40^{\circ}$) and the forward ($\theta < 35^{\circ}$)
regions, respectively.
A VD track element was defined by at least two $R\phi$ hits on different 
layers within a tolerance of 0.5 degrees.

\subsection{Event flavour identification}
\label{sec:idl}

The event flavour, in the $hadronic$ topologies with isolated leptons, 
was tagged by loosely identifying the leptons according to the criteria 
described in section \ref{sec:selec}:
in the h210 topology, events were tagged as electronic (muonic) events
if the final state lepton was identified as an electron (muon) and in
h220 if one of the final state leptons was identified as an electron
(muon) and the other one was not identified as a muon (electron).
In the tau channel, 
the momentum of the isolated lepton is expected to be lower than 
for the other leptonic flavours.
For the h220 topology the lower energy lepton is expected to be
the spectator lepton produced together with the excited one and 
in the tau channel it was
required that p$_\ell < 0.11~\sqrt{s}$. The same is true for the
h210 topology when it arises from the $W$ decay of a charged excited 
lepton ($\ell \ell^* \rightarrow \ell \nu W$).
In other cases, such as for neutral excited leptons, the final state lepton 
can be more energetic and in the tau channel it was required that 
p$_\ell < 0.22~\sqrt{s}$.

In the topologies corresponding to the pair production of neutral
heavy leptons, all events with more than two isolated leptons in the
final state were kept. If only two leptons were present, events were kept 
as candidates in the electron (muon) channel if both leptons were
identified as such.

The flavour identification for $leptonic$ topologies was performed using
the leptonic jet identification (see section~\ref{sec:selec}) 
and, whenever possible, the comparison 
between the measured momenta and the momenta computed using the rescaling
procedure (see section \ref{sec:lep2}). 

In the $leptonic$ topologies with no isolated photons, the rescaling
procedure was not applied, since there are always at least two neutrinos
involved. There is also at least one charged jet coming from the
decay of a $W$ or $Z$ boson and containing no relevant flavour information. 
Events were classified as electron or muon events whenever the lowest energy
jet was identified as such. All events were kept in the tau channel.    

In the topologies involving isolated photons, since there are no missing 
particles or there is only one particle lost along the beam pipe, the 
momenta can be computed imposing energy-momentum conservation.
Events were kept as candidates in the electron (muon) channel if
at least one of the jets was identified as an electron (muon) and no jets 
were identified as muons (electrons) and if $\chi^2_{charged}<5$. 
For the $\ell$101 topology, where background problems
are more severe, it was also required that $\chi^2_{photons}<5$.
Events were kept as candidates in the tau channel if $\chi^2_{charged}>5$ 
and $\chi^2_{photons}<5$. 
                           
\section{Results}
\label{sec:results}

The number of candidates at different selection levels are given in table  
\ref{tab:evhad} for the $hadronic$ topology, and in tables \ref{tab:evlep1} 
and \ref{tab:evlep2} for  the
$leptonic$ topology with (table \ref{tab:evlep1}) and without 
(table \ref{tab:evlep2}) isolated photons. 
Selection level 1 corresponds to the general criteria, 
before any specific topology cuts (section \ref{sec:selec}).
Level 2 corresponds to specific topology cuts, without flavour tagging
(sections \ref{sec:had} and \ref{sec:lep}). 
Flavour tagging is included in level 3 (section \ref{sec:idl}).
The numbers in brackets give the simulated SM background expectations.
The topologies marked with a P correspond to the pair production modes.
Whenever a topology is relevant for both the charged and the neutral new
leptons search and different selection criteria were applied 
(see sections~\ref{sec:hadsing} and \ref{sec:lep1}), the name of the topology
is followed by `char' or `neut'.
A given selection level is always a subsample of the previous one. 
The different flavours considered at a given level are
not exclusive.
In the different selection levels and topologies, fair agreement between
data and the SM expectation is found.

\begin{figure}[p]
\begin{center}
\vspace{-1cm}
\mbox{\epsfig{file=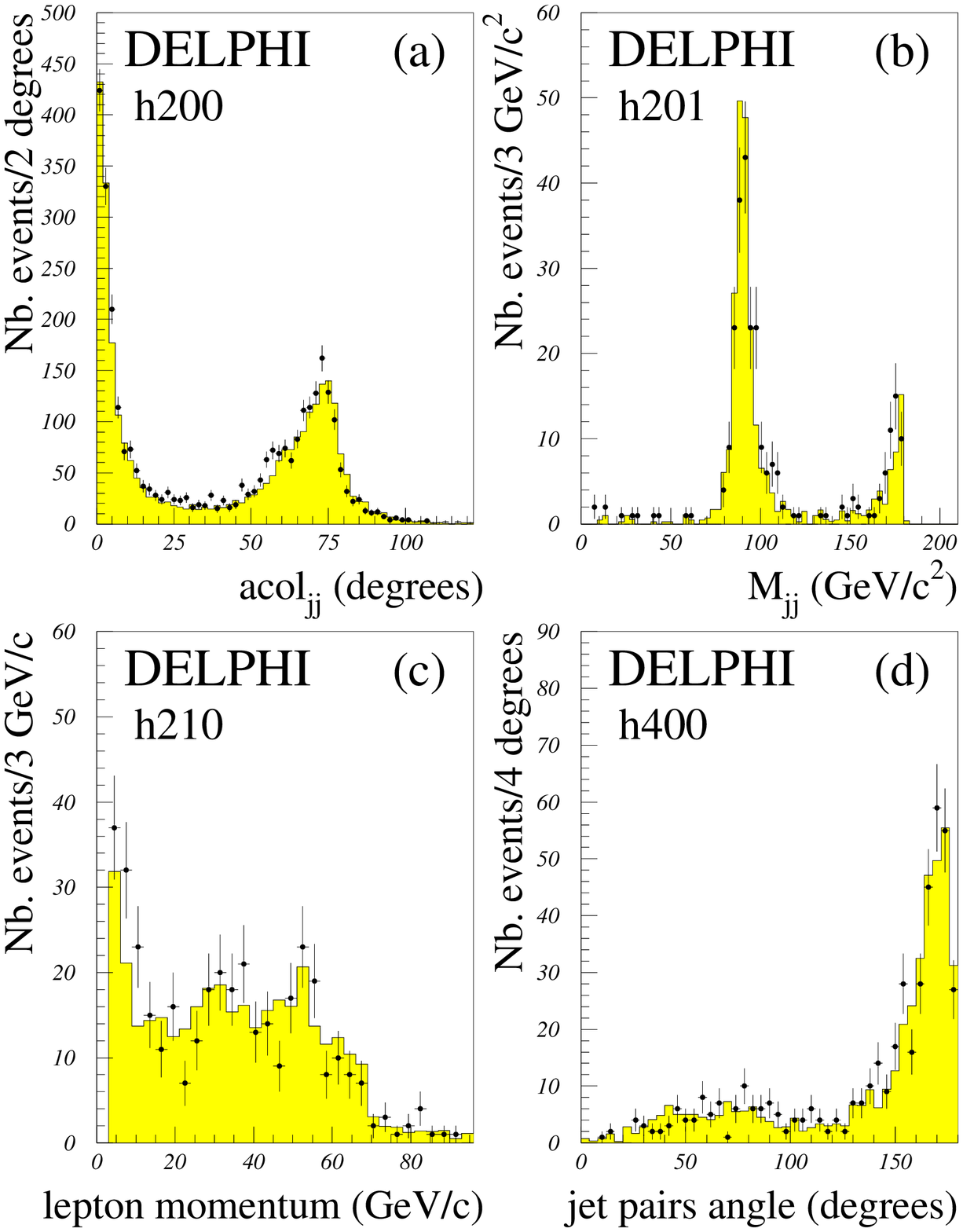,height=0.8\textheight}}
\vspace{-2mm}
\caption[]{Acolinearity {\bf(a)} and invariant mass {\bf(b)} of the two jets 
in the h200 and h201 topologies respectively, momentum of the lepton in 
the h210 topology {\bf(c)}, angle between the two jet pairs in the h400 
topology {\bf(d)} at 183~GeV.
The dots show the data and the shaded histograms show the SM simulation.}
\label{fig:had}
\end{center}
\end{figure}

\begin{figure}[p]
\begin{center}
\vspace{-1cm}
\mbox{\epsfig{file=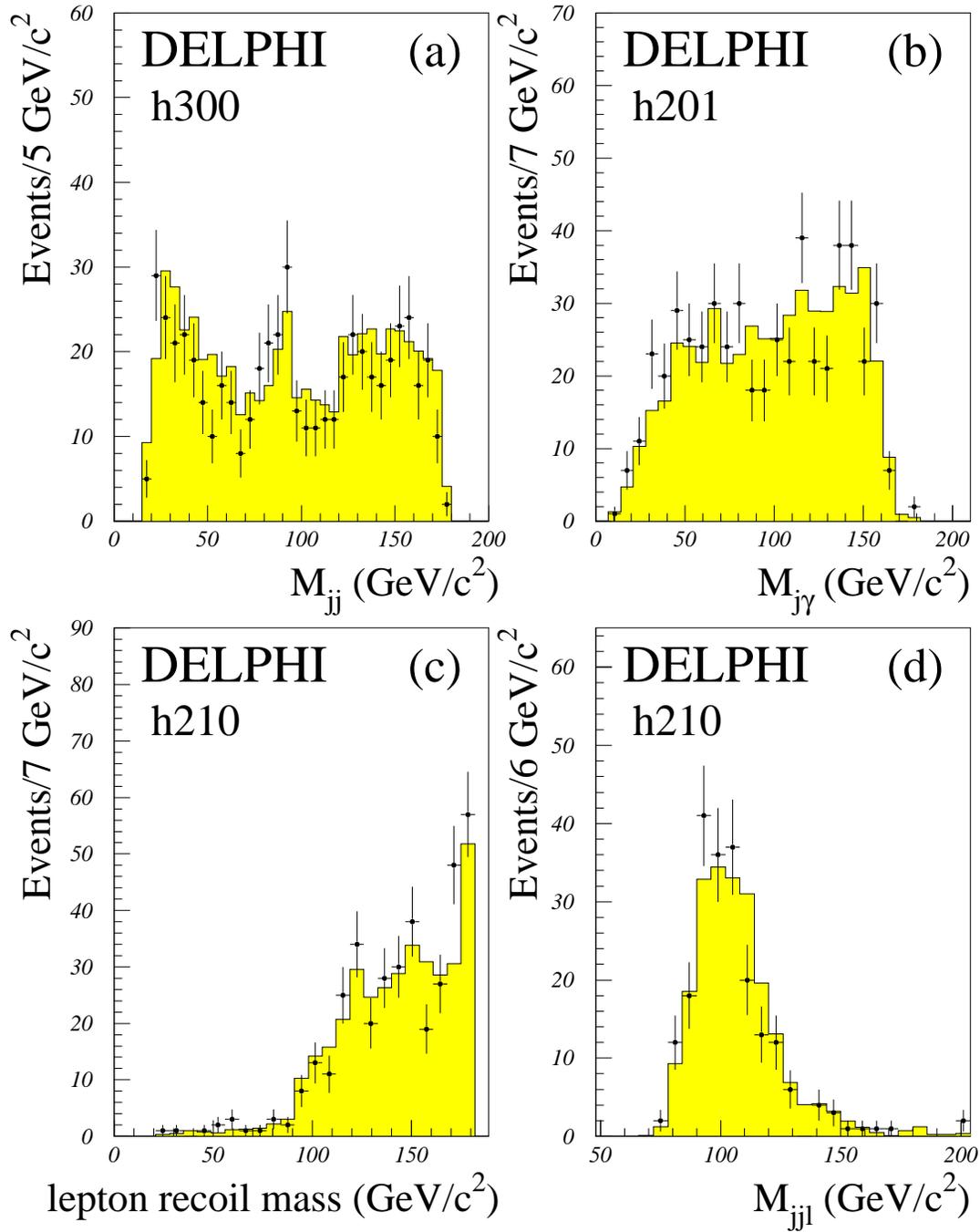,height=0.8\textheight}}
\vspace{-2mm}
\caption[]{Jet-jet {\bf(a)} and jet-photon {\bf(b)} invariant mass 
in the h300 and h201 topologies respectively, lepton recoil mass 
{\bf(c)} and jet-jet-lepton invariant mass {\bf(d)} in the h210 topology 
at 183~GeV.
The dots show the data and the shaded histograms show the SM simulation.}
\label{fig:had2}
\end{center}
\end{figure}

Data and SM simulation distributions at $\sqrt{s}=$183 GeV for the $hadronic$ 
topologies at selection level 1 are shown in figures \ref{fig:had} and
\ref{fig:had2}.
Figure \ref{fig:had}(a) and (b) show the jet-jet acollinearity 
and the jet-jet invariant mass for the h200 and h201 topologies respectively. 
In \ref{fig:had}(c) the momentum of the lepton in the h210 topology
is shown. Figure \ref{fig:had}(d) concerns the h400 topology and shows the 
angle between the two jet-pairs taken as $W$ candidates. 
There is, for all the distributions, a fair overall agreement.
As will be discussed below, 
the distributions shown in figure \ref{fig:had2} are the ones relevant for
signal mass reconstruction. The presence of a signal would correspond to
a peak in these variables. As before, there is reasonable overall
agreement and no relevant signal is observed. It should be noted that
in (a) and (b) there are three and two entries per event respectively
corresponding to the different possible jet-jet and jet-photon 
combinations.

\begin{figure}[p]
\begin{center}
\vspace{-1cm}
\mbox{\epsfig{file=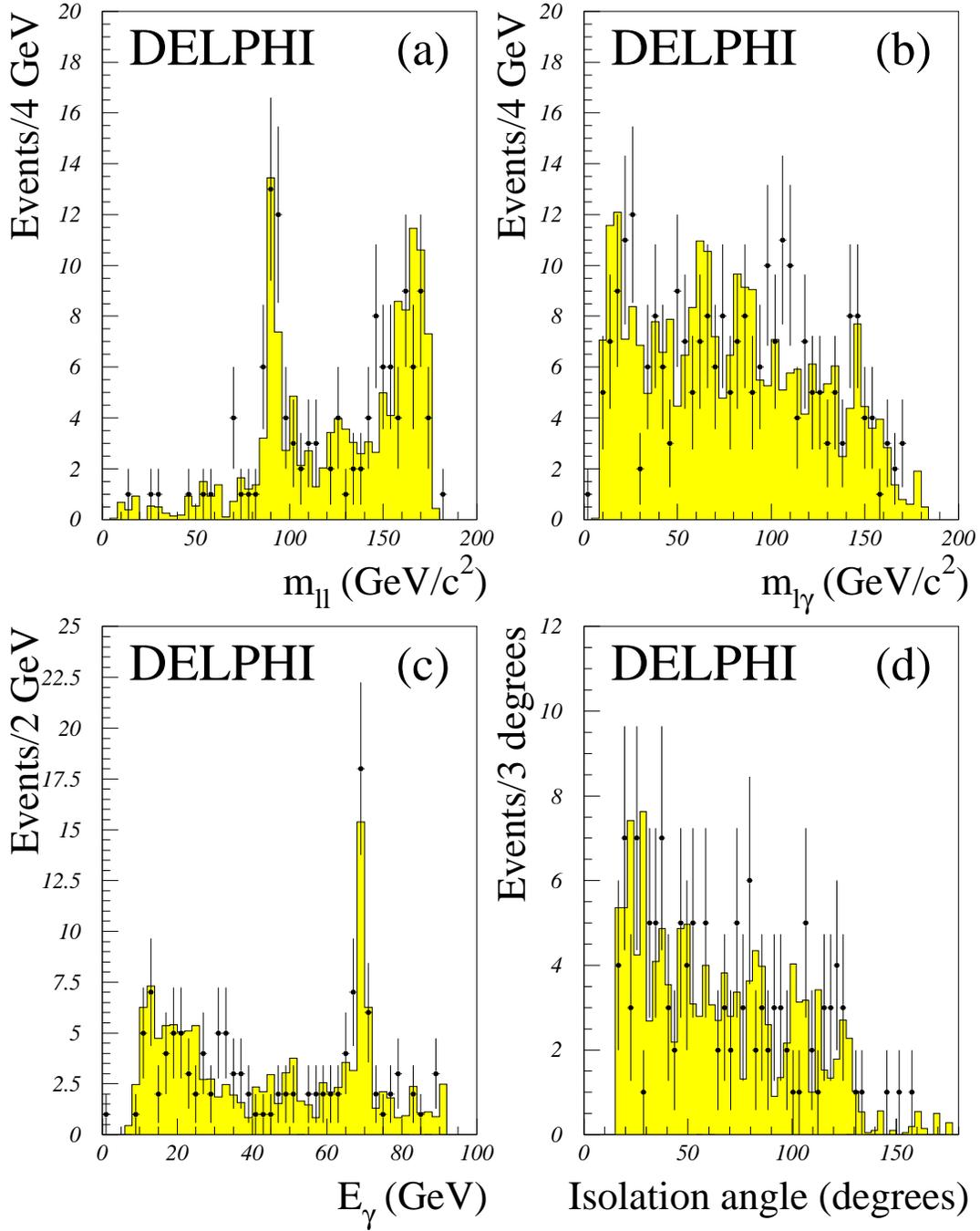,height=0.8\textheight}}
\vspace{-2mm}
\caption[]{Invariant mass of the two leptons {\bf(a)},
invariant mass of lepton-photon pairs {\bf(b)}, and
energy {\bf(c)} and isolation angle {\bf(d)} of the photon,
for the $\ell 201$ topology at 183~GeV.
The dots show the data and the shaded histograms show the SM simulation.}
\label{fig:lept}
\end{center}
\end{figure}

Figure \ref{fig:lept} shows distributions for the $\ell$201 topology
at the first level of the event selection for $\sqrt{s}=$183 GeV. 
Figure \ref{fig:lept}(a) shows the invariant mass for the
lepton-lepton pairs using the momenta calculated 
from the kinematic constraints.
The two possible $\ell \gamma$ invariant mass combinations are
plotted in figure~\ref{fig:lept}(b).
Figures \ref{fig:lept}(c) 
and~\ref{fig:lept}(d) display
the energy and
the isolation angle of the radiated photon.
There is reasonable agreement between the data and SM simulation.

In many topologies, the heavy fermion mass can be estimated from the momenta
and directions of final state particles. Relevant cases are the $\ell \gamma$
invariant mass for radiatively decaying excited leptons, the
jet-$\gamma$ and jet-jet invariant masses for excited quarks, 
the jet-jet-lepton invariant mass and the recoil mass of 
isolated leptons for the situations involving $W$ and $Z$ bosons. 
Signal simulation studies allowed the determination of the mass 
resolution for each situation.
In {\em leptonic} events, the mass resolution on the lepton-photon 
invariant mass, 
after applying the kinematic constraints, was found to be
about 1 GeV/$c^2$ for muons, 1.5 GeV/$c^2$ for electrons and 
2 GeV/$c^2$ for taus.
In the h300 and h201 topologies, the resolution on the jet-jet and 
jet-photon invariant masses after
the kinematic fits was found to be about 2~GeV/$c^2$.
For the h101 topology no kinematic fit was applied and the resolution
was around 20~GeV/$c^2$.
In {\em hadronic} events with isolated leptons, the resolution on the
lepton recoil mass (\mbox{$ m_{\ell}^2 = s - 2~k~P_\ell~ \sqrt{s}$}, 
where $k=1.0$ for electrons and muons and $k=1.4$ for taus to take in 
account the missing energy in the tau decay) is about 1 GeV/$c^2$  for muons, 
$3$ GeV/$c^2$ for electrons and 5 GeV/$c^2$ for taus. 
The resolution on the jet-jet-lepton invariant mass 
is about 5 GeV/$c^2$ for muons and 8 GeV/$c^2$ for electrons.
In this case, no mass reconstruction was attempted in the tau channel.

The results for single and double photon final state are as follows.
In the single photon channel,
three events having a single $\gamma$ in the barrel region 
(i.e. $\theta_{\gamma}=45^{\circ} - 135^{\circ}$) with $E_{\gamma} >80$ GeV
were found at $\sqrt{s}=183$ GeV, while 2.5 were expected
from the SM reaction $e^+e^- \rightarrow \gamma \nu \bar{\nu} $.
At $\sqrt{s}=172$  and $\sqrt{s}=161$ GeV, no events were found
with $E_{\gamma} >75$ GeV 
and $E_{\gamma} >70$ GeV respectively, 
while 0.02$\pm$0.01 and 0.08$\pm$0.03 were expected from the simulation. 


In the two photon channel 
four events with an acoplanarity greater than $10^{\circ}$ were found at
$\sqrt{s}=183$ GeV, while $ 0.4 \pm 0.1 $ events were expected from the QED
background reaction
$e^+e^- \rightarrow \gamma \gamma$ and $ 1.5 \pm 0.2 $ from the process
$e^+e^- \rightarrow Z \gamma \gamma$ with the $Z$ decaying into neutrinos.
Two events with an acoplanarity greater than $10^{\circ}$ were found at
$\sqrt{s}=172$ GeV, while 
$0.09 \pm 0.02 $ were expected from 
$e^+e^- \rightarrow \gamma \gamma$ and $0.61\pm0.02$ from 
$e^+e^- \rightarrow Z \gamma \gamma \rightarrow \nu \nu \gamma \gamma$. 
No candidates were found at $\sqrt{s}=$161 GeV while $0.8\pm0.1$ were
expected.

\section{Limits}
\label{sec:limits}
The search for the production of unstable heavy fermions
involves many final states. 
The relevance of the different final states depends, 
as discussed in section \ref{sec:theo3},
on the decay branching ratios which are a function of the heavy fermion
mass and of the coupling parameters.
\begin{table}[hbt]
\begin{center}
{
\vskip -0.4 cm
\begin{tabular}{|c|l|c|c|c|}
\hline
$\sqrt{s}$ & channel &$e$  &  $\mu$  & $\tau$ 
\\
\hline
\hline

    &
$\ell^* \rightarrow \ell\gamma$& 44(53$\pm$4) & 17(16$\pm$1) & 21(22$\pm$2) 
\\ \cline{2-5}
    &
$\ell^* \rightarrow \nu W$ & 17(16$\pm$2) & 7(8$\pm$1)& 22(17$\pm$2)   
\\ \cline{2-5}
    &
$\ell^* \rightarrow \ell Z$ & 26(31$\pm$2) & 22(17$\pm$1) & 46(41$\pm$3)
\\ 
\cline{2-5}
\cline{2-5}
183 & $\nu^* \rightarrow \nu \gamma$ & \multicolumn{3}{c|}{3(2.5$\pm$0.3)}
\\ \cline{2-5}
GeV &
$\nu^* \rightarrow \ell W$ & 17(18$\pm$2) & 10(12$\pm$2) & 26(25$\pm$2)   
\\ \cline{2-5}
&$\nu^* \rightarrow \nu Z $ & \multicolumn{3}{c|}{15(12$\pm$1)} 
\\ 
\cline{2-5}
\cline{2-5}
 &$q^* \rightarrow q \gamma$ & \multicolumn{3}{c|}{120(114$\pm$5)} \\ 
\cline{2-5}
 &$q^* \rightarrow q g$ & \multicolumn{3}{c|}{84(98$\pm$5)} \\ 
\hline
\hline
    &
$\ell^* \rightarrow \ell\gamma$& 8(9$\pm$1) & 7(4.8$\pm$0.4) & 5(5.2$\pm$0.8) 
\\ \cline{2-5}
    &
$\ell^* \rightarrow \nu W$ & 2(3.6$\pm$0.5) & 1(1.9$\pm$0.3)& 1(4.2$\pm$0.4)   
\\ \cline{2-5}
    &
$\ell^* \rightarrow \ell Z$ & 4(4.9$\pm$0.5) & 3(2.5$\pm$0.3) & 7(7.3$\pm$0.6)
\\ 
\cline{2-5}
\cline{2-5}
172 & $\nu^* \rightarrow \nu \gamma$ & \multicolumn{3}{c|}{0(0.02$\pm$0.01)}
\\ \cline{2-5}
GeV &
$\nu^* \rightarrow \ell W$ & 4(3.7$\pm$0.4) & 3(2.1$\pm$0.4) & 7(5.5$\pm$0.5)   
\\ \cline{2-5}
&$\nu^* \rightarrow \nu Z $ & \multicolumn{3}{c|}{3(2.7$\pm$0.4)}
\\ 
\cline{2-5}
\cline{2-5}
 &$q^* \rightarrow q \gamma$ & \multicolumn{3}{c|}{36(33$\pm$2)} \\ 
\cline{2-5}
 &$q^* \rightarrow q g$ & \multicolumn{3}{c|}{27(24$\pm$1)} \\ 
\hline
\hline
    &
$\ell^* \rightarrow \ell\gamma$& 5(15$\pm$1) & 6(7$\pm$1) & 10(7$\pm$1) 
\\ \cline{2-5}
    &
$\ell^* \rightarrow \nu W$ & 2(3.2$\pm$0.4) & 4(1.8$\pm$0.2)& 4(2.9$\pm$0.3)   
\\ \cline{2-5}
    &
$\ell^* \rightarrow \ell Z$ & 1(2.9$\pm$0.4) & 2(1.2$\pm$0.2) & 7(4.0$\pm$0.4)
\\ 
\cline{2-5}
\cline{2-5}
161 & $\nu^* \rightarrow \nu \gamma$ & \multicolumn{3}{c|}{-}
\\ \cline{2-5}
GeV &
$\nu^* \rightarrow \ell W$ & 1(3.2$\pm$0.4) & 2(1.3$\pm$0.2) & 7(4.5$\pm$0.5)   
\\ \cline{2-5}
&$\nu^* \rightarrow \nu Z $ & \multicolumn{3}{c|}{2(1.4$\pm$)0.3}
\\ 
\cline{2-5}
\cline{2-5}
 &$q^* \rightarrow q \gamma$ & \multicolumn{3}{c|}{40(44$\pm$3)} \\ 
\cline{2-5}
 &$q^* \rightarrow q g$ & \multicolumn{3}{c|}{29(30$\pm$2)} \\ 
\hline
\end{tabular}
}
\end{center}
\vskip -0.4 cm
\caption[]{Number of excited fermion candidates for the different decay channels
and the different centre-of-mass energies in the single production mode.
The numbers in brackets correspond to the simulated SM background expectations.}
\label{tab:cand1}
\end{table}
\begin{table}[hbt]
\begin{center}
{
\begin{tabular}{|c|l|c|c|c|}
\hline
$\sqrt{s}$ & channel & $e$ & $\mu$  & $\tau$  \\
\hline
\hline
 & $\ell^* \rightarrow \ell\gamma$ 
& 0(0.2$\pm$0.1) & 0(0.2$\pm$0.1 ) & 1(0.9$\pm$0.2) 
\\ 
\cline{2-5}
183 & $\nu^* \rightarrow \nu\gamma$   & \multicolumn{3}{c|}{4(1.8$\pm$0.2)}  
\\ 
\cline{2-5}
GeV & $\ell^*,L^{\pm},E_i^{\pm} \rightarrow \nu W$ 
& \multicolumn{3}{c|}{22(18$\pm$1)} 
\\ 
\cline{2-5}
 & $\nu^*,L^0,E_i^0 \rightarrow \ell W$  
& 5(2.6$\pm$0.5) & 5(2.9$\pm$0.5) & 17(13$\pm$1)      
\\ 
\hline
\hline
 & $\ell^* \rightarrow \ell\gamma$ 
&0(0.04$\pm$0.04)& 0(0.04$\pm$0.04) & 0(0.1$\pm$0.1) 
\\ 
\cline{2-5}
172 & $\nu^* \rightarrow \nu\gamma$   & \multicolumn{3}{c|}{2(0.8$\pm$0.1)}  
\\ 
\cline{2-5}
GeV & $\ell^*,L^{\pm},E_i^{\pm} \rightarrow \nu W$ 
& \multicolumn{3}{c|}{5(4$\pm$0.4)}
\\ 
\cline{2-5}
 & $\nu^*,L^0,E_i^0 \rightarrow \ell W$  & 
1(0.2$\pm$0.1) & 2(0.4$\pm$0.1) & 4(2.4$\pm$0.3)      
\\ 
\hline
\hline
 & $\ell^* \rightarrow \ell\gamma$ 
& 1(0.04$\pm$0.04) & 0(0.1$\pm$0.1) & 1(0.3$\pm$0.1) 
\\ 
\cline{2-5}
161 & $\nu^* \rightarrow \nu\gamma$   & \multicolumn{3}{c|}{0(0.8$\pm$0.1)}  
\\ 
\cline{2-5}
GeV & $\ell^*,L^{\pm},E_i^{\pm} \rightarrow \nu W$ 
& \multicolumn{3}{c|}{3(3.5$\pm$0.5)}
\\ 
\cline{2-5}
 & $\nu^*,L^0,E_i^0 \rightarrow \ell W$  
& 0(0.5$\pm$0.2) & 0(0.4$\pm$0.1) & 1(2.0$\pm$0.3)      
\\ 
\hline
\end{tabular}
}
\end{center}
\caption[]{Number of heavy lepton candidates for the different decay channels
and the different centre-of-mass energies in the pair production modes. 
The numbers in brackets correspond to the simulated SM background expectations.}
\label{tab:cand2}
\end{table}
\begin{table}[hbt]
\begin{center}
{
\begin{tabular}{|l|c|c|c|}
\hline
 channel &  $e$   &   $\mu$   &  $\tau$ 
\\
\hline
\hline

$\ell^* \rightarrow \ell\gamma$& 33  & 59  & 33  
\\ \hline
$\ell^* \rightarrow \nu W$  & 21 & 43 & 32     
\\ \hline
$\ell^* \rightarrow \ell Z$ & 26 & 53 & 23 
\\ 
\hline
\hline
$\nu^* \rightarrow \nu \gamma$ &  \multicolumn{3}{c|}{42} 
\\ \hline
$\nu^* \rightarrow \ell W$ & 44 & 48 & 17    
\\ \hline
$\nu^* \rightarrow \nu Z $ &  \multicolumn{3}{c|}{21} 
\\ \hline\hline
  $q^* \rightarrow q \gamma$ & \multicolumn{3}{c|}{31} \\ \hline
 $q^* \rightarrow q g$      & \multicolumn{3}{c|}{19} \\ \hline
\end{tabular}

\vspace{0.4 cm}

\begin{tabular}{|l|c|c|c|}
\hline
channel & $e$ & $\mu$  & $\tau$  \\
\hline
\hline
$\ell^* \rightarrow \ell\gamma$ & 28 & 37 & 26  
\\ \hline
$\nu^* \rightarrow \nu\gamma$   &  \multicolumn{3}{c|}{46} 
\\ 
\hline
\hline
$\ell^*,L^{\pm},E_i^{\pm} \rightarrow \nu W $  &  \multicolumn{3}{c|}{14}      
\\ \hline
$\nu^*,L^0,E_i^0 \rightarrow \ell W $ & 27   & 37  & 20    
\\ \hline
\end{tabular}
}
\end{center}
\caption[]{ Efficiencies (in percentage) 
for the single (upper) and pair (lower) 
production modes at a centre-of-mass energy of 183 GeV. 
The values were obtained with excited lepton masses of 
170 GeV/$c^2$ and 80 GeV/$c^2$ for single and pair production
respectively.} 
\label{tab:eff}
\end{table}
\begin{table}[htb]
\begin{center}
{
\begin{tabular}{|c|c|c|c|} \hline
$L^{\pm}$ & $L^0 \rightarrow e W$ & 
$L^0 \rightarrow \mu W$ & $L^0 \rightarrow \tau W$ 
\\ 
\hline
\hline
78.3 & 76.5 & 79.5 & 60.5   \\ 
\hline
\end{tabular}

\vspace{0.4 cm}

\begin{tabular}{|c||c|c|c|c|} \hline
        &  $E_i $  & $N_{e}$ & $N_{\mu}$ & $N_{\tau}$ \\
\hline
\hline
vector  & 81.3  & 87.3 & 88.0 & 81.0 \\
\hline
mirror  & 78.3 & 76.5 & 79.5 & 60.5 \\
\hline
\end{tabular}

\vspace{0.4 cm}

\begin{tabular}{|c||c|c|c|} \hline
         & $e^*$      &     $\mu^*$
&       $\tau^*$     \\ 
\hline
\hline
$f=f^{'}$     & 90.7  & 90.7 & 89.7 \\ \hline
$ f= - f^{'}$ & 81.3  & 81.3 & 81.3 \\ \hline 
\end{tabular}
\begin{tabular}{|c||c|c|c|} \hline
    & $\nu^*_{e}$    &   
      $\nu^*_{\mu}$  &     $\nu^*_{\tau}$ \\
\hline
\hline
$f=f^{'}$     & 
87.3 & 88.0 & 81.0 \\ \hline
$ f= - f^{'}$ & 
90.0 & 90.0 & 90.0 \\ \hline
\end{tabular}
}
\end{center}
\caption[]{ {Lower limits (in GeV/$c^2$) at 95 \% CL on the 
unstable heavy lepton masses from the pair production modes.
Starting from the top, 
the three tables correspond to sequential, non-canonical and
excited leptons respectively.}}
\label{tab:limits}
\end{table}

The numbers of excited fermion candidates in the single production 
topologies, as well as the SM
expectations, are summarized in table \ref{tab:cand1} 
for the different excited fermion types and decay modes and for the three
centre-of-mass energies.
It should be noted that these numbers result from the combination of the
different topologies (tables~\ref{tab:evhad}, \ref{tab:evlep1} and 
\ref{tab:evlep2}) and there is, in many cases, overlap between the candidates
selected in the different decay channels listed in table~\ref{tab:cand1}.

For exotic leptons only pair production was considered. 
The number of heavy lepton candidates found and the SM simulation 
expectations 
at $\sqrt{s}$=183, 
172 and 161 GeV are summarized in table \ref{tab:cand2}
for pair production modes.

The possible heavy fermion masses can be deduced in many of the topologies, 
as referred to in the previous section.
Events for which the mass could not be estimated were treated as
candidates for all possible mass values. 
 
The efficiencies, including the trigger efficiency, are given in table 
\ref{tab:eff} for all the studied channels and for chosen heavy fermion mass 
values at $\sqrt{s}=$~183~GeV.
The efficiency levels are very similar for scaled masses at the other 
centre-of-mass energies.
The trigger efficiency was estimated to be greater than 85\% and 95\% 
for the single and the double photon channels respectively. 
For all the other topologies it is greater then 99\% (being essentially
100\% for all hadronic topologies). 
The dependence of the efficiency on the mass is weak, due to the
combination of the several topologies and of the different
centre-of-mass energies.

The limits were computed using the method described in \cite{read}.
This is the method used in the DELPHI Higgs search analysis, and is well suited
both for the combination of channels and for the inclusion of  
mass information. Each topology at each centre-of-mass energy 
was treated as a channel, and the used mass resolution depends on the
specific reconstruction procedure for each topology,
as explained in the previous section. 

\begin{figure}[p]
\begin{center}
\vspace{-1cm}
\mbox{\epsfig{file=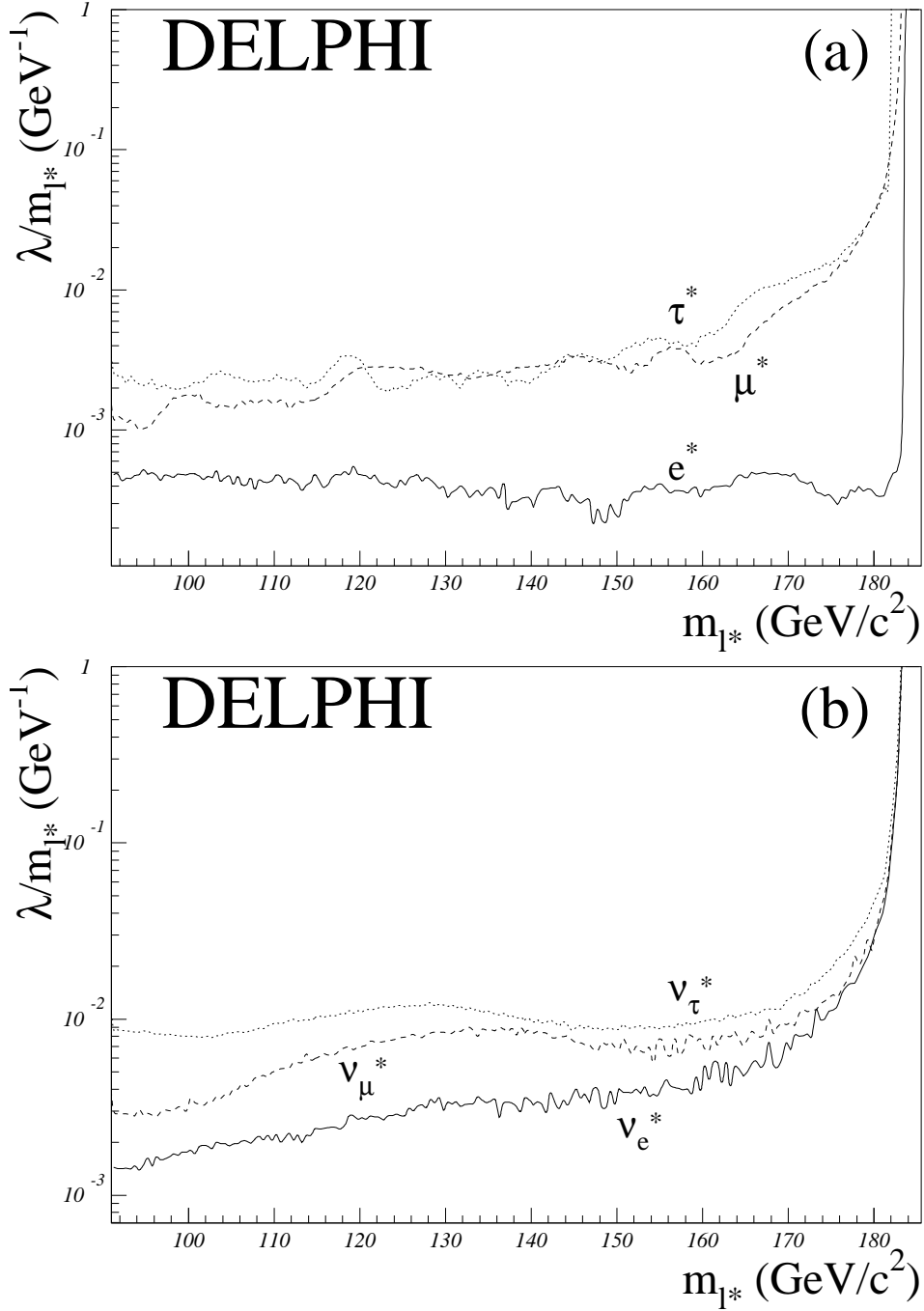,height=0.8\textheight}}
\vspace{-2mm}
\caption[]
{ Results on single production of excited charged {\bf(a)} and neutral {\bf(b)}
leptons assuming $f=+f^{'}$. The lines show the 
upper limits at 95\% CL
on the ratio $\lambda/m_{\ell^*}$ between the coupling of the excited  
lepton and its mass as a function the 
mass.}
\label{fig:lim1}
\end{center}
\end{figure}

\begin{figure}[p]
\begin{center}
\vspace{-1cm}
\mbox{\epsfig{file=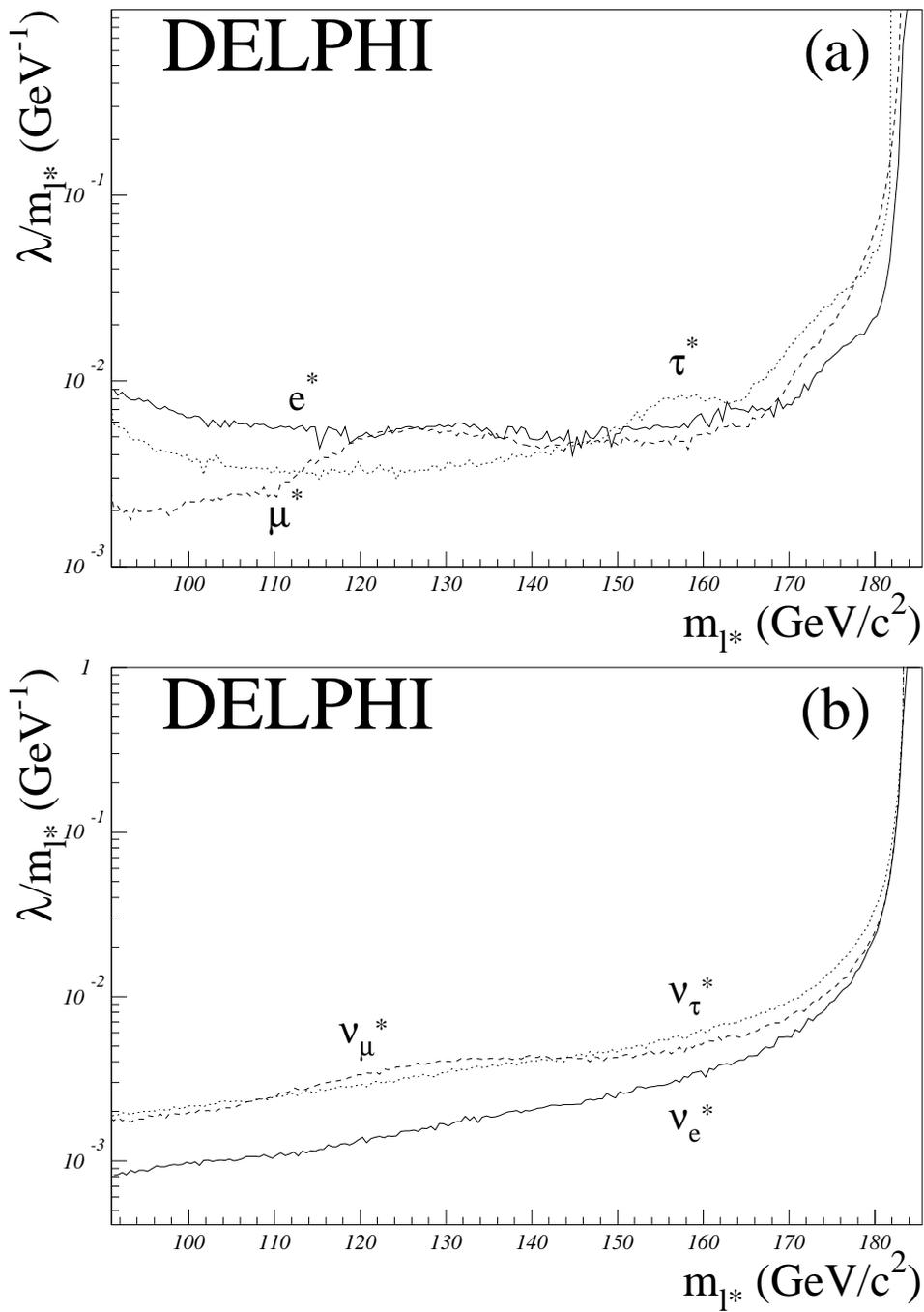,height=0.8\textheight}}
\vspace{-2mm}
\caption[]{ As figure~4, but for  $f=-f^{'}$.}
\label{fig:lim2}
\end{center}
\end{figure}

For the single production of excited fermions,
the production cross section
is a function not only of the mass of the particle but also of the
ratio of the coupling of the excited lepton to its mass. 
95\% confidence level (CL) upper limits 
on the ratio $\lambda/m_{f^*}$ (see section \ref{sec:theo3})
as a function of the $f^*$ mass were derived. 
Figures~\ref{fig:lim1}  and \ref{fig:lim2} show these limits for the excited
leptons assuming $f=f^{'}$ and $f=-f^{'}$ respectively.

\begin{figure}[p]
\begin{center}
\vspace{-1cm}
\mbox{\epsfig{file=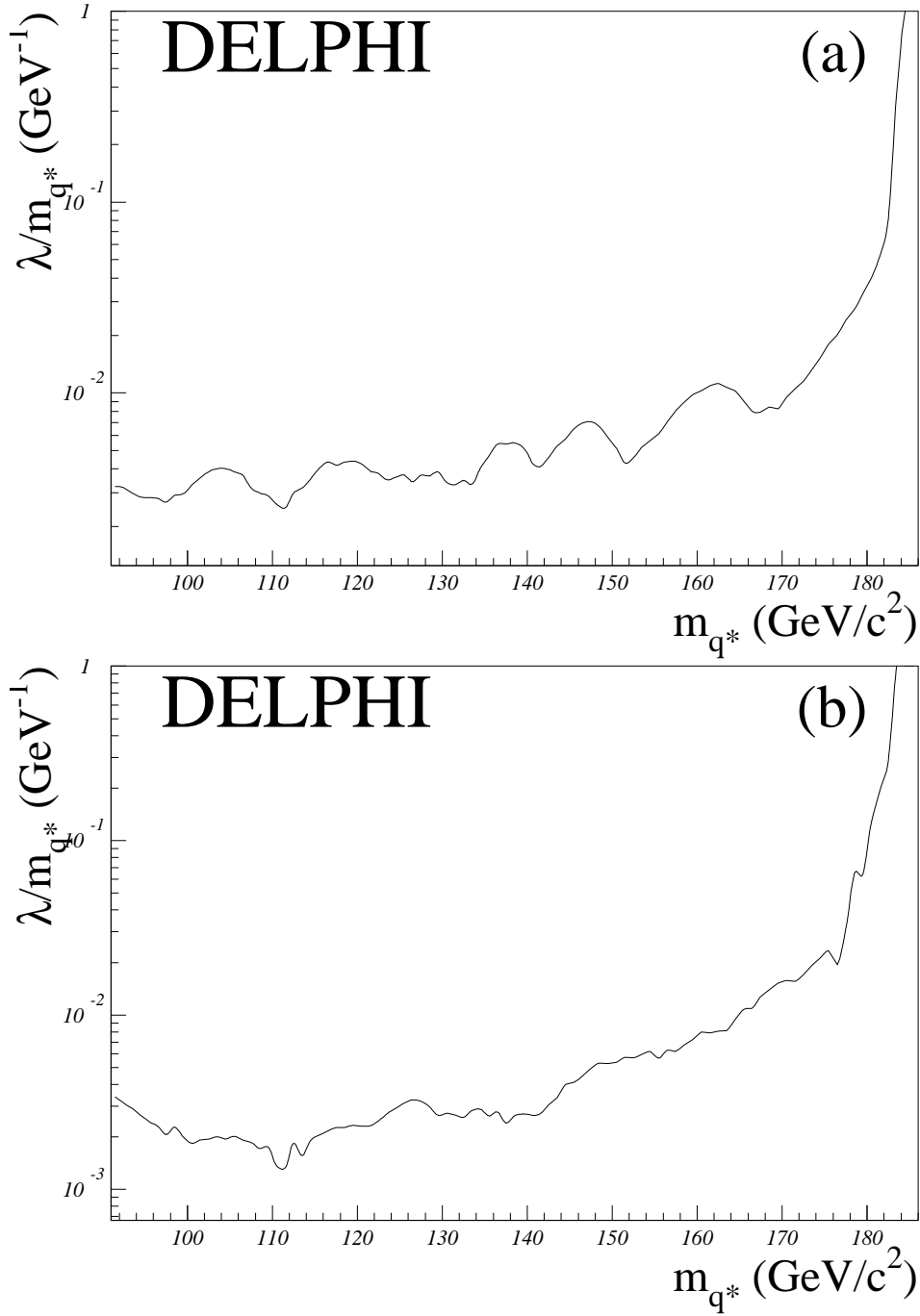,height=0.8\textheight}}
\vspace{-2mm}
\caption[]
{ Results on single production of excited quarks assuming a branching
ratio of 100\%
for the photon {\bf(a)} or the gluon {\bf(b)} decay modes.
The lines show the upper limits at 95\% CL
on the ratio $\lambda/m_{q^*}$ between the coupling of the excited quark
and its mass as a function of the
mass.}
\label{fig:lim_qx}
\end{center}
\end{figure}

Figures~\ref{fig:lim_qx}(a) and \ref{fig:lim_qx}(b) show the limits 
on the single production of excited quarks, namely limits 
on $\lambda/m_{q^*}$ multiplied by the branching ratio of the $q^*$ into
q$\gamma$ and into q$g$ respectively.
These limits were obtained assuming up-type excited quarks.
For down-type excited quarks the cross section limits are about 15\% higher 
in the studied mass region due to the lower expected production cross-section.

The lower limits at 95\% CL on the masses of 
pair-produced unstable heavy leptons are given in table \ref{tab:limits}.
In the excited leptons case, limits are given for both $f=f^{'}$ and $f=-f^{'}$.
In the case of the sequential leptons, decays into each of the leptonic 
families are considered. 

\begin{figure}[p]
\begin{center}
\vspace{-1cm}
\mbox{\epsfig{file=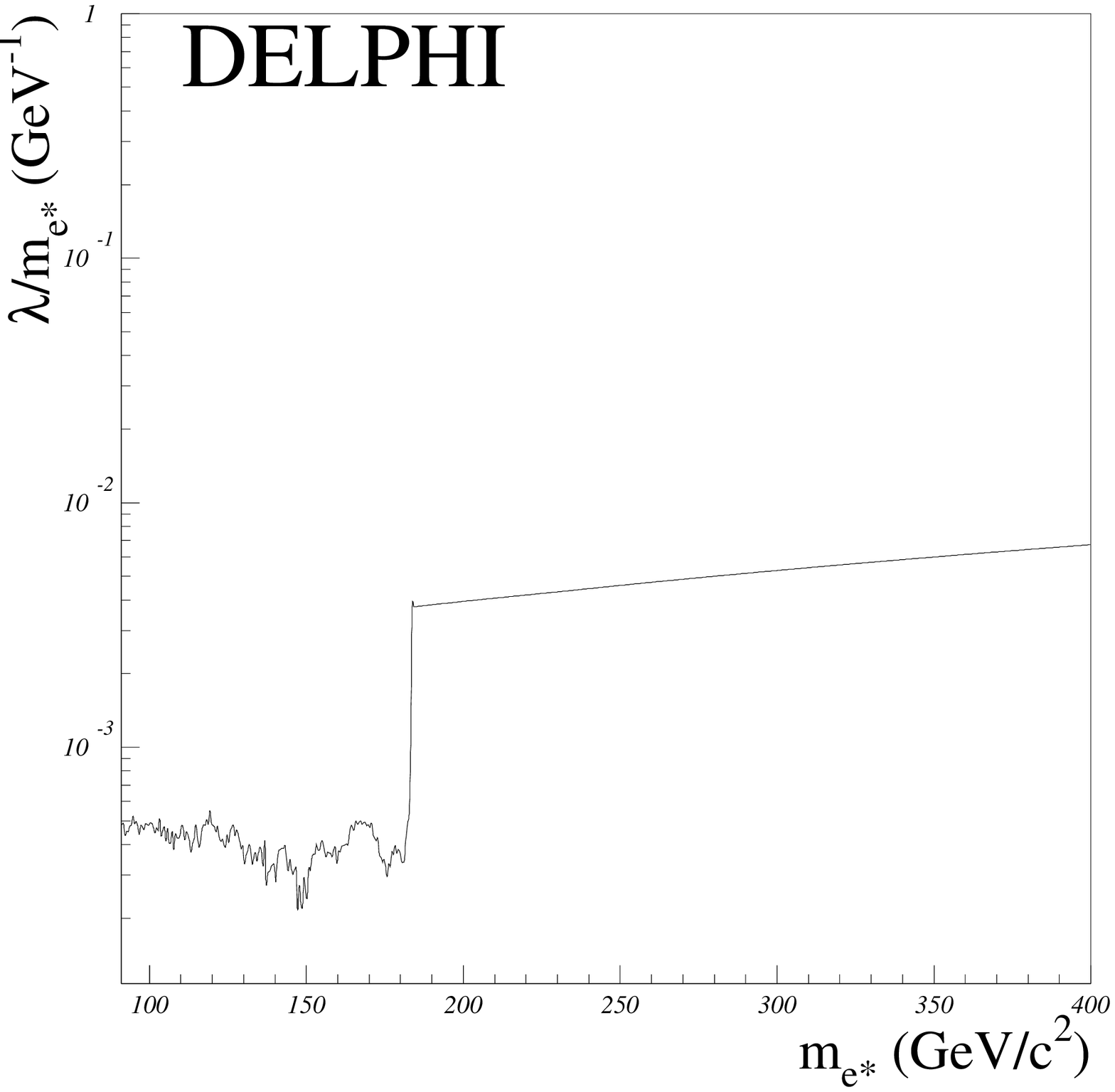,height=0.6\textheight}}
\vspace{-2mm}
\caption[]
{Combined excited electron limits for $f=f^{'}$ from direct and indirect 
searches. The line shows the upper limits at 95\% CL
on the ratio $\lambda/m_{e^*}$ between the coupling of the excited  
electron and its mass as a function of the mass.
Up to the kinematic limit the result is dominated by the single 
production direct search. Above this value the limit it is the one coming from
the indirect search using $e^+ e^- \rightarrow \gamma \gamma$.}
\label{fig:clim}
\end{center}
\end{figure}

Figure~\ref{fig:clim} shows the limit on the excited electron production
for $f=f^{'}$ obtained by combining the result of the direct search
(figure~\ref{fig:lim1}(a)) with the indirect result from the search for
deviations in the $e^+ e^- \rightarrow \gamma \gamma (\gamma)$ 
differential cross-section~\cite{photons}. 
We can thus extend the limit to regions
above the kinematic limit.
 
\section{Conclusions}

DELPHI data corresponding to integrated luminosities of 47.7 pb$^{-1}$, 10
pb$^{-1}$ and 10 pb$^{-1}$ at the centre-of-mass energies of 183 GeV, 172 GeV 
and 161 GeV respectively have been analysed.
A search for unstable heavy fermions 
decaying promptly through
$\gamma$, $Z$, $W$ or gluon emission
was performed. 
No significant signal was observed.

The search for pair production of heavy leptons resulted in
the mass limits quoted in table \ref{tab:limits}.
The search for single production of excited fermions gave the limits on the 
ratio $\lambda/m_{f}^{*}$ shown in Figures~\ref{fig:lim1}, \ref{fig:lim2},
\ref{fig:lim_qx} and \ref{fig:clim}.
These results considerably extend
the limits recently set from the runs of LEP at centre-of-mass
energies of 161 GeV and 172 GeV or previously at 
LEP1 and HERA \cite{delphi1,delphi2,others,delphip3}.  
 
\newpage
\subsection*{Acknowledgements}
\vskip 3 mm
We would like to thank Jorge Rom\~ao for the very useful discussions.

 We are greatly indebted to our technical 
collaborators, to the members of the CERN-SL Division for the excellent 
performance of the LEP collider, and to the funding agencies for their
support in building and operating the DELPHI detector.\\
We acknowledge in particular the support of \\
Austrian Federal Ministry of Science and Traffics, GZ 616.364/2-III/2a/98, \\
FNRS--FWO, Belgium,  \\
FINEP, CNPq, CAPES, FUJB and FAPERJ, Brazil, \\
Czech Ministry of Industry and Trade, GA CR 202/96/0450 and GA AVCR A1010521,\\
Danish Natural Research Council, \\
Commission of the European Communities (DG XII), \\
Direction des Sciences de la Mati$\grave{\mbox{\rm e}}$re, CEA, France, \\
Bundesministerium f$\ddot{\mbox{\rm u}}$r Bildung, Wissenschaft, Forschung 
und Technologie, Germany,\\
General Secretariat for Research and Technology, Greece, \\
National Science Foundation (NWO) and Foundation for Research on Matter (FOM),
The Netherlands, \\
Norwegian Research Council,  \\
State Committee for Scientific Research, Poland, 2P03B06015, 2P03B03311 and
SPUB/P03/178/98, \\
JNICT--Junta Nacional de Investiga\c{c}\~{a}o Cient\'{\i}fica 
e Tecnol$\acute{\mbox{\rm o}}$gica, Portugal, \\
Vedecka grantova agentura MS SR, Slovakia, Nr. 95/5195/134, \\
Ministry of Science and Technology of the Republic of Slovenia, \\
CICYT, Spain, AEN96--1661 and AEN96-1681,  \\
The Swedish Natural Science Research Council,      \\
Particle Physics and Astronomy Research Council, UK, \\
Department of Energy, USA, DE--FG02--94ER40817. \\

\newpage

\newpage
                                                    
\newpage

\newpage

\newpage

\begin{thebibliography}{99}
\bibitem{read}
A.L. Read, 
``Optimal statistical analysis of search results based 
on the likelihood ratio and
its application to the search for the MSM Higgs 
boson at $\sqrt{s}$=161 and 172 GeV", DELPHI note 97-158 PHYS 737.
\end{thebibliography}
\end{document}